\documentclass[%
 reprint,
 amsmath,amssymb,
 aps,
]{revtex4-2}
 
\usepackage{graphicx}
\usepackage{dcolumn}
\usepackage{bm}

\usepackage[font=small,labelfont=bf]{caption}
\usepackage{subcaption}
\usepackage{wrapfig}
\usepackage{hyperref}
\usepackage{comment}
\usepackage{bbm}
\usepackage{nicefrac}
\usepackage[utf8]{inputenc}
\usepackage{textgreek}
\usepackage{graphicx}
\usepackage{booktabs}
\usepackage{multirow}

\def\beq{\begin{equation}\displaystyle}
\def\eeq{\end{equation}}
\def\bea{\begin{eqnarray}\displaystyle}
\def\eea{\end{eqnarray}}

\def\bry{\begin{array}}
\def\ery{\end{array}}

\begin{document}

\preprint{APS/123-QED}

\title[Sample title]{Covariant Contrastive Learning for Uncertainty-Aware Anomaly Detection}

\author{Shelley Tong}
\email{tshelley@mit.edu}
 \author{Philip Harris}
 \email{pcharris@mit.edu}
 \author{Gaia Grosso}%
 \email{gaiag795@mit.edu}
\affiliation{ 
NSF AI Institute for Artificial Intelligence and Fundamental Interactions, Cambridge, MA\\
MIT Laboratory for Nuclear Science, Cambridge, MA
}%
\date{\today}

\begin{abstract}
Machine-learning-based anomaly detection (AD) offers a promising, model-agnostic alternative to traditional LHC analyses, allowing to search for many signals at once. Recent advances in representation learning motivate the use of neural embeddings to map high-dimensional physics observables into low-dimensional latent spaces better suited to statistical inference. However, the propagation of systematic uncertainty in embedded spaces remains poorly understood, severely limiting the application of these methods to real LHC analyses.
We address this gap with a machine learning strategy that uses a likelihood-based regularization to improve the structure of embedded spaces. Building on previous work using supervised contrastive learning for physics-aware embeddings, our approach trains them jointly with a downstream parameterized classification task that incorporates continuous uncertainties as nuisance parameters. This method returns a latent space that covaries predictably with systematic shifts, and a model of such distortions that can be exploited in the downstream statistical test for anomaly detection.
Using simulated CMS Level-1 trigger data, we show that our method successfully models continuous uncertainties in a 4D latent space with a linear parametric downstream classifier, improving both the interpretability and robustness of the statistical anomaly detection task. This framework offers a general way to study and control systematic uncertainties in latent spaces, opening the way to a new scalable statistical analysis workflow for anomaly detection at the LHC, and other high energy physics experiments.
\end{abstract}

\maketitle


\section{Introduction}
\label{sec:intro}

The Large Hadron Collider (LHC) experiments have performed several tests of the Standard Model (SM) of particle physics, and continue to produce measurements of increasingly high precision. However, they have yet to find any convincing evidence of new physics. One reason may be methodological: most searches are optimized for a specific signal hypothesis and lose sensitivity to signatures that fall outside it. Since the space of possible new-physics signatures is far larger than the set of models that can be tested individually, dedicated searches can only ever cover a small fraction. This has motivated the development of \emph{signal-agnostic} searches based on anomaly detection (AD), which leverages machine learning to identify deviations between the observed data and the SM expectation. By searching for deviations without assuming a particular signal model, a single analysis can be sensitive to a wide range of possible new-physics signatures~\cite{ATLAS:2020iwa,ATLAS:2023azi,ATLAS:2025obc, CMS:2024nsz, CMS:2025sch}.

\begin{figure*}[t]
    \centering
    \includegraphics[width=1\linewidth]{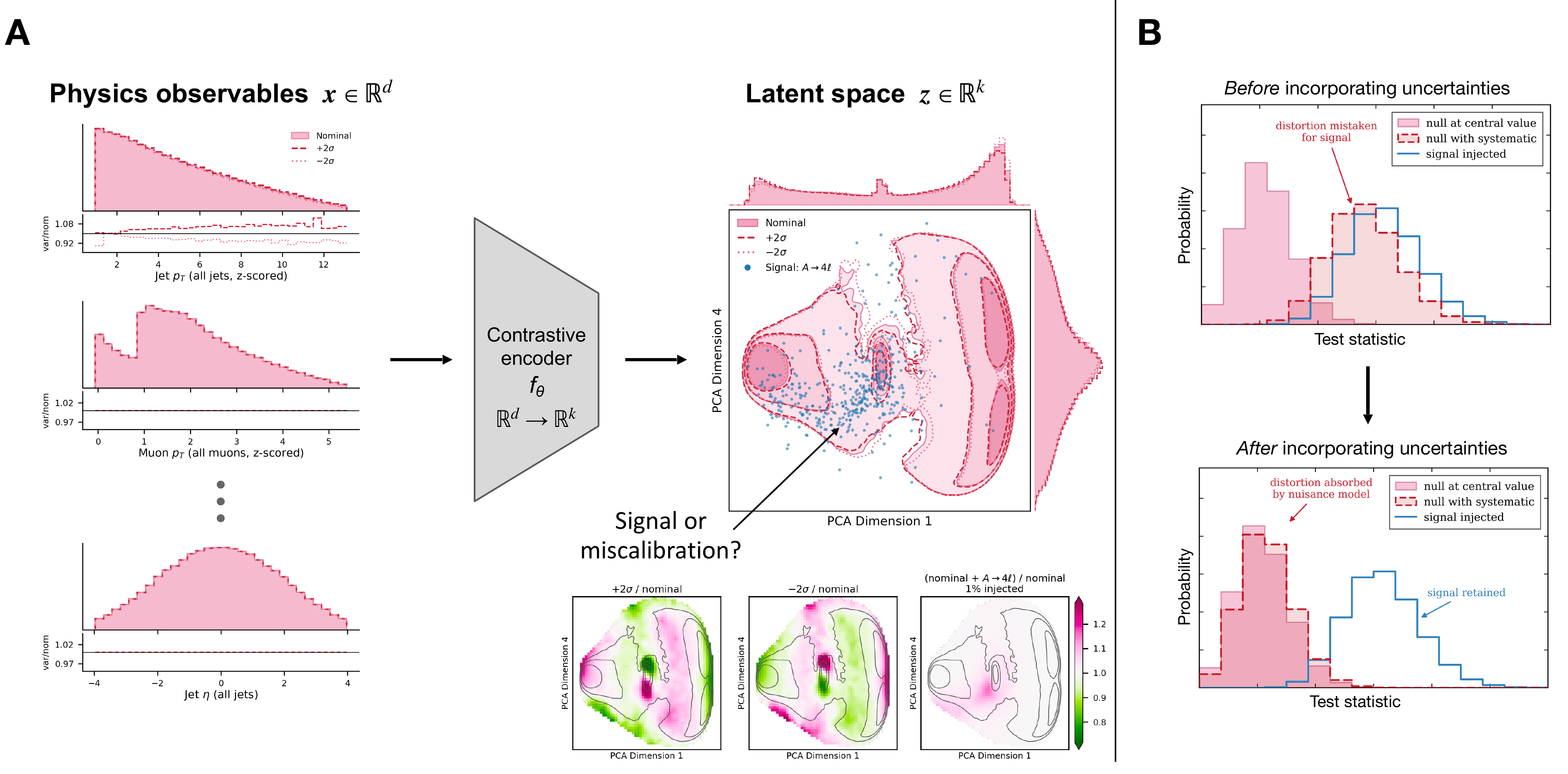}
\caption{\textbf{Calibration enables robust detection.} \textbf{(A)} We map a $d$-dimensional space of physics observables into a $k$-dimensional contrastive latent space to perform anomaly detection. The ratio plots at the bottom show that $\pm2\sigma$ distortions of the nominal (signal-free) data distribution (left and center) produce latent-space deformations resembling those of a genuine signal, such as the 1\% $A\to4\ell$ injection shown in the rightmost panel, and can therefore be mistaken for anomalous signal if not properly modeled. \textbf{(B)} We illustrate the risks of false detection in the absence of a rigorous calibration: systematic effects altering the data distribution trigger an AD test to fire as if a real signal was present (top). Incorporating uncertainties in the AD test makes its response invariant to systematic distortions, recovering a correct false positive rate (bottom).}
    \label{fig:intro}
\end{figure*}

The proton-proton collision events collected and analyzed by the LHC experiments are characterized by very high-dimensional representations, and the most viable strategy to robustly perform statistical analysis is to compress the representation into a low-dimensional summary statistic, and perform statistical inference in the reduced space. While for a signal-specific search, one knows the best way to reduce the data to one or a few summary statistics thanks to physics knowledge and supervised machine learning algorithms, in signal-agnostic approaches, the target is not known a priori, and the risk of detection failure due to a wrong summarization becomes critical. Supervised contrastive learning has emerged as an effective way to build such a summary for AD, organizing events by physics process while preserving
the structure that separates background from potential anomalies~\cite{Metzger:2025ecl,bright2025autoscidact,Li:2026azw,Cheng:2026iiu}.
Compressions of this kind mitigate the look-elsewhere effect inherent in high-dimensional searches and simultaneously reduce the complexity of the problem, retaining power in the downstream statistical test and making it scalable to the full rate of LHC data (see panel A of Figure~\ref{fig:intro}).

For any of this to translate into a complete LHC result, however, systematic uncertainties must be propagated rigorously through the entire pipeline. A credible discovery claim requires a calibrated and robust statistical test. At LHC experiments, every source of uncertainty is treated as a nuisance parameter and profiled. This is exactly where current latent-space AD methods
fall short. A nonlinear encoder can turn a simple, smooth systematic shift of the observables into an arbitrarily complicated deformation of the latent space. The dependence of the background density on the nuisance parameters can then become difficult to model, breaking down the profile-likelihood machinery and any robustness guarantee with it. Without robustness guarantees, the test can mistake a
systematic shift for a signal leading to an increased chance of false discovery. Panel B of Figure~\ref{fig:intro} illustrates the problem: when systematic uncertainties are inadequately modeled in the test, the resulting shifts in the data can trigger false alarms (dashed red histogram) that mimic genuine new physics signals (blue step histograms). Treating systematic uncertainties properly and rigorously in the embedding space is, therefore, not just an additional refinement, but a central challenge that must be addressed before these methods can be used in real analyses.

In this work we propose to address this gap by introducing a likelihood-based regularization to jointly train a contrastive encoder so that its latent space covaries \emph{linearly} with the nuisance parameters. This restores a tractable, analytic model of the background density ratio in the
latent space, which we feed into the New Physics Learning Machine (NPLM) profile-likelihood test~\cite{d2019learning,DAgnolo:2019vbw,dAgnolo:2021aun}. The result is an AD workflow that
is simultaneously low-dimensional and statistically rigorous: systematic distortions are
absorbed by the nuisance model rather than faked as signal. Using simulated CMS Level-1
trigger data, we demonstrate that the method models a realistic jet-energy-scale uncertainty
in a 4D latent space, passes closure tests on nuisance-shifted backgrounds, and retains
sensitivity to a range of injected beyond-the-SM (BSM) signals.

The rest of this work is organized as follows. Section~\ref{sec:related} reviews related
work, covering contrastive learning for anomaly detection at the LHC, equivariant
contrastive learning, and the profile-likelihood treatment of systematic uncertainties on
which we build. Section~\ref{sec:method} presents our method: a likelihood-based regularization
that fine-tunes a contrastive encoder so that its latent space covaries linearly
with the nuisance parameters, together with the resulting profile-likelihood
anomaly-detection test. Section~\ref{sec:experiments} validates the approach on
simulated CMS Level-1 trigger data, reporting both closure tests on
nuisance-shifted backgrounds and sensitivity to a range of injected BSM signals.
Section~\ref{sec:conclusion} summarizes our findings and outlines
directions for future work.

\section{Related Work}
\label{sec:related}

\subsection{Contrastive learning for anomaly detection}
\label{sec:related-cl}

Contrastive learning constructs a latent representation by pulling together the representations of similar examples and pushing apart dissimilar ones. Choosing a latent space with fewer dimensions than the input space also allows this representation to compress the inputs and perform dimensionality reduction. Since the introduction of instance-discrimination frameworks such as
SimCLR~\cite{chen2020simclr}, it has become a standard tool for representation learning. In
high energy physics it was first deployed in a self-supervised form to build symmetry-aware
jet representations: JetCLR~\cite{dillon2022symmetries} maps jet constituents into a space
that is approximately invariant to physically motivated augmentations, later it was used in the context of augmenting simulations to mitigate systematics~\cite{Harris:2024sra}, and closely related
self-supervised approaches have since been applied directly to model-agnostic anomaly
detection~\cite{dillon2022selfsupervised}. Building on this line of work, Metzger et
al.~\cite{Metzger:2025ecl} and related studies~\cite{bright2025autoscidact} established that
\emph{supervised} contrastive losses with physics-process labels produce latent spaces
well suited for signal-agnostic AD at the LHC, with domain-knowledge-informed label encoding
being the key ingredient for statistical discovery performance. Our encoder is built with
exactly this supervised contrastive strategy. None of these works, however, address how
systematic uncertainties propagate through the learned embedding once it is used for
inference---the gap we fill here.

\subsection{Equivariant contrastive learning}
\label{sec:related-ecl}

Controlling \emph{how} a learned representation responds to a transformation of its input is
precisely the subject of equivariant representation learning. Standard contrastive learning
enforces \emph{invariance}, training representations to be unchanged under a chosen set of
augmentations. Invariance is a special case of the more general notion of \emph{equivariance}, in which a transformation of the input induces a corresponding, specified transformation of its representation. Dangovski et al.~\cite{dangovski2021equivariant} added an auxiliary prediction task to self-supervised learning based on invariance: identifying the transformation applied to each input. This encourages broken equivariance for some transformations while retaining invariance for others. They showed that this approach improves the semantic quality of the learned representations. We adapt this idea to the physics setting: the transformations
of interest are not geometric augmentations but the systematic distortions of the
observables parameterized by nuisance parameters $\nu$, and the property we require is not
invariance but a controlled, \emph{linear} covariance of the latent density with $\nu$. Rather than requiring invariance under these distortions, we require the latent density to vary with $\nu$ in a controlled, linear manner (cf. Eq.~\ref{eq:logr-poly}). Enforcing this equivariance-like structure is what renders the downstream profile-likelihood test tractable, as detailed in Section~\ref{sec:method}.

\subsection{Dealing with systematic uncertainties in statistical anomaly detection}
\label{sec:nplm}

The statistical framework in which we ultimately deploy the embedding is the ``New Physics Learning Machine'' (NPLM)~\cite{d2019learning,DAgnolo:2019vbw}, a statistical test for signal-agnostic detection based on the likelihood-ratio, which includes a treatment of systematic uncertainties using nuisance parameters. Since our work is directly built on this framework, we review it below.

\paragraph{The NPLM hypothesis test.}
The NPLM method~\cite{d2019learning,DAgnolo:2019vbw}
treats signal-agnostic anomaly detection as a likelihood-ratio test between an
an empirical model of the observed dataset $\mathcal{D}=\{x_i\}$ and the Standard Model (SM) hypothesis represented by a reference set $\mathcal{R}$. In the language of hypothesis testing, $R$ is simply the null hypothesis---the Standard Model prediction in the absence of new physics---which,in NPLM terminology, we call the \emph{Reference} and denote by $R$ throughout. Under the alternative hypothesis $H_\zeta$, the data density distribution is defined as a local reweighting of the Reference distribution by the exponential of a neural network $h_\zeta$ with trainable parameters $\zeta \in \Omega$,
\begin{equation}
  n(x|H_\zeta) = e^{\,h_\zeta(x)}\, n(x|R).
  \label{eq:nplm-alt}
\end{equation}
The test statistic is the maximum log-likelihood-ratio
\begin{equation}
  t(\mathcal{D}) = 2\,\max_{\zeta\in \Omega}\,
    \log\frac{\mathcal{L}(H_\zeta|\mathcal{D})}{\mathcal{L}(R|\mathcal{D})},
  \label{eq:nplm-t}
\end{equation}
obtained directly from the minimum of a training loss acting on
$\mathcal{D}$ and $\mathcal{R}$. Because $h_\zeta$ is a flexible, unbiased
approximant, the test is sensitive to generic departures from the SM rather than
to a single signal model. 
Ref.~\cite{DAgnolo:2019vbw} showed that, under the Reference hypothesis and with a suitable regularization scheme for the test network $h_\zeta$, $t$ closely follows a $\chi^2$ distribution.

\paragraph{Continuous systematics as nuisance parameters.}
The treatment above assumes the Reference is known exactly, which is never the case:
detector calibration, theory, and PDF uncertainties all distort the expected SM
distribution, and these effects must be propagated into any LHC result for it to be
trusted. Ref.~\cite{dAgnolo:2021aun} extends NPLM to incorporate them following the
canonical profile-likelihood treatment, by associating each source of uncertainty with a
nuisance parameter $\nu$ constrained by an auxiliary likelihood $\mathcal{L}_{\rm aux}(\nu)$.
The Reference hypothesis becomes composite, $R_\nu$, and the alternative gains the same
nuisance dependence, $H_{\nu,\zeta}$:

\begin{equation}
  t(\mathcal{D}) = 2\log\frac{\max_{\nu,\,\zeta}\, [{\cal L}(H_{\nu,\zeta}\,|\,\mathcal{D})\cdot{\cal L}_{\rm aux}(\nu)]}{\max_{\nu'}\, [{\cal L}(R_{\nu'}\,|\,\mathcal{D})\cdot{\cal L}_{\rm aux}(\nu')]},
  \label{eq:teststat}
\end{equation}
Normalizing both likelihoods to that of the central-value Reference $R_{\nu_0}$, where $\nu_0$ is the best prior estimate of the nuisance parameter, allows us to express the test statistic as the difference of two positive-definite terms:
\begin{equation}
  t(\mathcal{D},\mathcal{A}) =
    \tau(\mathcal{D},\mathcal{A}) - \Delta(\mathcal{D},\mathcal{A}),
  \label{eq:t-decomp}
\end{equation}
and the nuisances $\nu$,
\begin{align}
  \tau   &= 2\,\max_{\nu,\zeta}\,
    \log\!\left[\frac{\mathcal{L}(H_{\nu,\zeta}|\mathcal{D})}{\mathcal{L}(R_{\nu_0}|\mathcal{D})}
      \cdot\frac{\mathcal{L}_{\rm aux}(\nu)}{\mathcal{L}_{\rm aux}(\nu_0)}\right],
    \label{eq:tau}\\
  \Delta &= 2\,\max_{\nu}\,
    \log\!\left[\frac{\mathcal{L}(R_{\nu}|\mathcal{D})}{\mathcal{L}(R_{\nu_0}|\mathcal{D})}
      \cdot\frac{\mathcal{L}_{\rm aux}(\nu)}{\mathcal{L}_{\rm aux}(\nu_0)}\right].
    \label{eq:Delta}
\end{align}
Intuitively, $\tau$ is the full ``discovery'' term: it measures how much better
the flexible alternative $H_{\nu,\zeta}$ with nuisances describes the data than the central-value Reference. As a result, we see the test statistic deviates from $R_{\nu_0}$ for \emph{any} reason, whether it is a genuine signal or a mismodeled systematic. The correction $\Delta$ contains no test network: it measures how much of that same departure can be absorbed by the Reference alone, simply by pulling the nuisance $\nu$ away from its central value. Subtracting $\Delta$ from $\tau$ therefore removes the part of the discrepancy that is explainable as a systematic distortion, so that $t=\tau-\Delta$ isolates genuinely anomalous structure. Both terms are maximum-likelihood log-ratios and are asymptotically $\chi^2$-distributed; the cancellation between them is precisely what keeps $P(t|R_\nu)$ close to a $\nu$-independent $\chi^2$, so that a single calibration on $R_{\nu_0}$ yields a $p$-value valid across the allowed range of $\nu$.
\begin{figure*}[t]
    \setlength{\textfloatsep}{2pt}
    \centering
    \includegraphics[width=\linewidth]{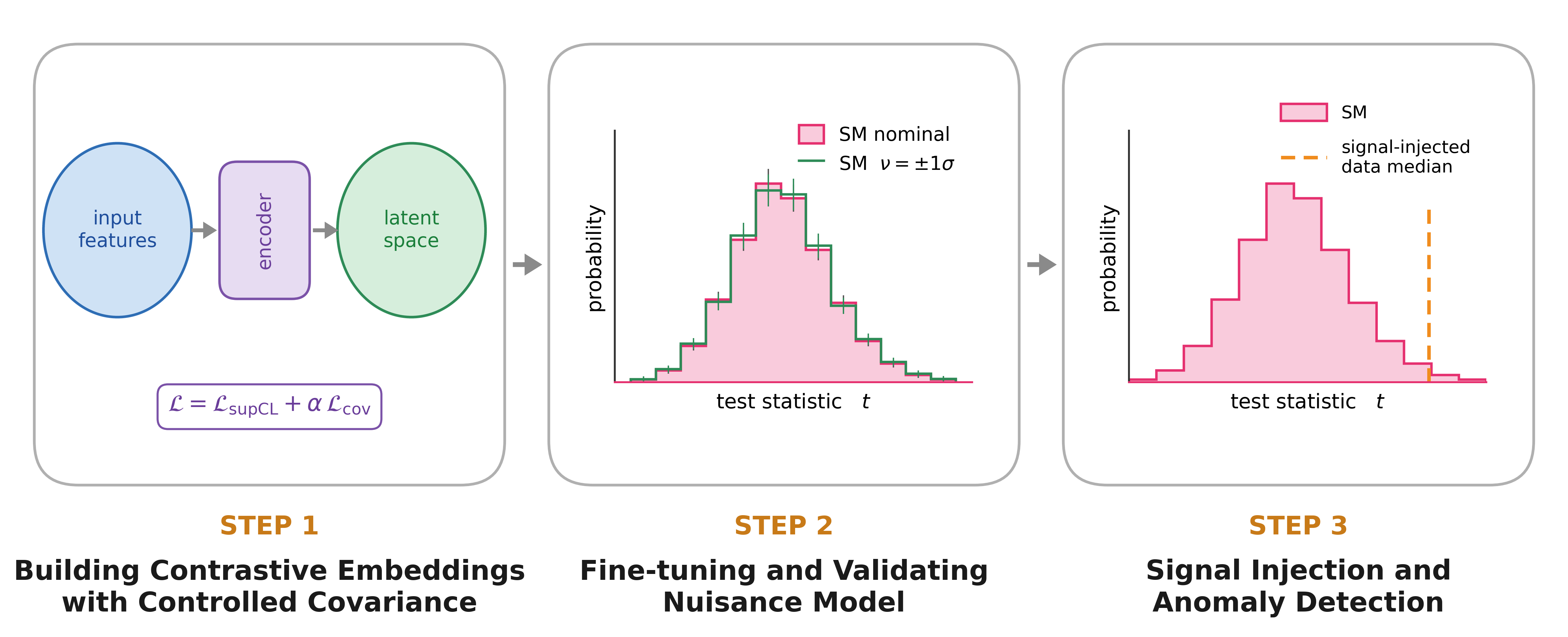}
    \caption{\textbf{Overview of the three-stage end-to-end workflow.} (Step 1) A contrastive encoder compresses high-dimensional physics observables into a low-dimensional latent space, trained jointly with a regularization that makes the effect of the nuisance parameters $\nu$ on the latent distributions linear. (Step 2) The learned linear nuisance model is fine-tuned and validated with closure tests in the New Physics Learning Machine (NPLM) framework. (Step 3) Signal is injected and detected as a shift of the NPLM test-statistic distribution.}
    \label{fig:FlowChart}
\end{figure*}
\paragraph{Modeling the nuisance dependence.}
Both $\tau$ and $\Delta$ require the density ratio between the shifted and
nominal Reference distributions,
\begin{equation}
  r(x|\nu) = \frac{n(x|R_\nu)}{n(x|R_{\nu_0})},
\end{equation}
to be known as a function of $\nu$ at every point $x$. Ref.~\cite{dAgnolo:2021aun}
models its logarithm with a low-order polynomial in $\nu$,
\begin{equation}
  \log r(x|\nu)
    = \nu\,\delta_1(x) + \tfrac{1}{2}\nu^2\,\delta_2(x) + \dots,
  \label{eq:logr-poly}
\end{equation}
whose coefficient functions $\delta_a(x)$ are learned by training a parameterized classifier between the nominal and shifted samples with a quadratic (mean-squared) loss. This expansion is tractable only when the systematic effect is a small, smooth deformation of the distribution---a condition that holds in the space of physical observables, and is not guaranteed to survive a nonlinear encoder. Related strategies for handling systematics in learned classifiers include parameterized networks~\cite{baldi2016parameterized}, uncertainty-aware inference~\cite{de2019inferno}, and invariance via adversarial decorrelation~\cite{louppe2017learning,kasieczka2020disco}.

\paragraph{Linear Nuisance Dependence.}
The methods we build on were designed to act on physical observables, where the
small-deformation condition behind Eq.~\eqref{eq:logr-poly} may or may not hold depending on the problem under study, limiting the application of the method. Our goal is move the statistical test from the space of the physical observables to the latent space of a contrastive encoder $f_\theta$, that we have the freedom to train or fine-tune. A nuisance shift $x(\nu_0)\mapsto x(\nu)$ induces a transformation $z(\nu_0)\mapsto z(\nu)$ of the embedding. The latter could be large and difficult to approximate with a polynomial expansion. The polynomial model of $\log r(z|\nu)$ then breaks down resulting in a poor estimation of the correction term $\Delta$ (Eq.~\eqref{eq:Delta}). We resolve it by training the encoder so that the latent log-density-ratio is, by construction, \emph{linear} in $\nu$. This recovers a controlled $\Delta$ term and an interpretable, profilable nuisance model directly in the embedded space.

\section{Method}
\label{sec:method}

Our end-to-end analysis workflow includes three stages, as illustrated in Figure~\ref{fig:FlowChart}:
\begin{enumerate}
  \item \textbf{Building Contrastive Embeddings with Controlled Covariance.} We map high-dimensional
    physics observables into a low-dimensional latent space with supervised contrastive learning,
    training the encoder jointly with a likelihood-based regularization. The resulting embedding
    preserves the physics structure needed for anomaly detection while forcing continuous systematic
    uncertainties to act on the latent space in a controlled way, such that their effect is
    \emph{linear} in the nuisance parameter $\nu$.
  \item \textbf{Fine-tuning and Validating the Linear Nuisance Model.} We freeze the embedding, further
    fine-tune the linear model of the nuisance dependence, and validate it with NPLM closure tests
    performed in the presence of systematic uncertainties.
  \item \textbf{Signal Injection and Anomaly Detection.} We inject signal and run the profile-likelihood
    NPLM test directly in the latent space, quantifying the resulting shift of the test-statistic
    distribution and converting it into a calibrated significance.
\end{enumerate}
We describe the three stages in turn below; their empirical implementation and
validation are reported in Section~\ref{sec:experiments}.

\subsection{Stage 1: Building neural embeddings with a controlled covariance}
\label{sec:stage1}
Supervised contrastive learning offers a scalable route to anomaly detection by
performing dimensionality reduction through neural embeddings: it compresses physics
observables into a low-dimensional latent space while preserving sensitivity to statistical anomalies. Refs.~\cite{Metzger:2025ecl,bright2025autoscidact}
showed that contrastive losses defined with physics-process labels yield latent spaces
well suited to signal-agnostic anomaly detection at the LHC, with domain-informed label
encoding being the key ingredient for statistical discovery performance.

We train a transformer-based encoder $f_\theta:\mathbb{R}^d\rightarrow\mathbb{R}^k$
($k\ll d$) with the supervised contrastive loss~\cite{khosla2020supervised},
\begin{equation}
  \mathcal{L}_\mathrm{supCL} = -\sum_{i} \frac{1}{|P(i)|}
  \sum_{p \in P(i)} \log
  \frac{e^{\mathbf{z}_i \cdot \mathbf{z}_p / \tau}}
       {\sum_{a\neq i} e^{\mathbf{z}_i \cdot \mathbf{z}_a / \tau}},
\end{equation}
where $\mathbf{z}_i = f_\theta(\mathbf{x}_i)/\|f_\theta(\mathbf{x}_i)\|$ is the
$\ell_2$-normalized embedding, $P(i)$ indexes the events in the batch produced by the
same background physics process as event $i$, and $\tau$ is a hyperparameter denoted temperature. With labels taken from simulation, this organizes the latent space by
background physics process, clustering SM events of the same type. 
This loss does not account for systematic uncertainties. To include systematic uncertainties, we add an additional term to the loss that controls systematic variations on the latent space.

A continuous systematic uncertainty is described by a nuisance parameter $\nu$ that
induces a deterministic distortion of the observables, $\mathbf{x}(\nu_0)\mapsto\mathbf{x}(\nu)$. After encoding, this becomes a transformation
of the latent distribution, $\mathbf{z}(\nu_0)\mapsto\mathbf{z}(\nu)$. Without any
constraint on $f_\theta$, this transformation can be arbitrarily nonlinear in $\nu$,
making the latent log-density-ratio
\begin{equation}
\label{eq:ratio}
  r(\mathbf{z}|\nu) = \frac{n(\mathbf{z}|R_\nu)}{n(\mathbf{z}|R_{\nu_0})}
\end{equation}
difficult to model. As discussed in Section~\ref{sec:nplm}, the NPLM strategy requires uncertainties to be well approximated by a polynomial model, so recovering a well behaved transformation in the embedded space is essential.

To this end, we impose a \emph{linear covariance} condition on the latent space. Under this penalty loss the encoder is induced to organize the embedding so that the log-density-ratio is linear in $\nu$,
\begin{equation}
  \label{eq:linear}
  \log r(\mathbf{z}|\nu)
  = \log \frac{n(\mathbf{z}|R_\nu)}{n(\mathbf{z}|R_{\nu_0})}
  = g_\phi(\mathbf{z}) \cdot \nu,
\end{equation}
where $g_\phi(\mathbf{z})$ depends on the embedding alone. Under this condition $\nu$ is
directly estimable by maximum likelihood in the latent space, and the nuisance profiling
required by the statistical test becomes tractable.

We enforce condition~\eqref{eq:linear} by training the encoder jointly with a model
$g_\phi$ of the log-density-ratio. This mirrors equivariant self-supervised
learning~\cite{dangovski2021equivariant}, where an auxiliary head that predicts the
applied transformation improves representation quality; here we replace the geometric
transformations with systematic effects parameterized by $\nu$. Concretely, we
generate labeled pairs $(\mathbf{x}(\nu_0),\,\mathbf{x}(\nu))$ on a discrete grid of
$\nu$ spanning the expected range $[-\sigma,+\sigma]$, and train $g_\phi$ as a
parameterized classifier between the nominal and shifted distributions
following~\cite{dAgnolo:2021aun}, minimizing the quadratic loss
\begin{equation}
  \mathcal{L}_\mathrm{cov} =
    \frac{1}{N_\nu}\sum_{\nu}\left[
      \sum_{\mathbf{z} \sim R_{\nu_0}}\sigma\!\left(g_\phi(\mathbf{z})\cdot\nu\right)^2
    + \sum_{\mathbf{z} \sim R_{\nu}} \!\left[1 - \sigma\!\left(g_\phi(\mathbf{z})\cdot\nu\right)\right]^2
    \right],
  \label{eq:lcov}
\end{equation}
where $\sigma$ denotes the sigmoid function and the sum runs over the grid of nuisance
values. By the likelihood-ratio argument of~\cite{dAgnolo:2021aun}, the minimizer
satisfies $g_\phi(\mathbf{z})\cdot\nu \to \log r(\mathbf{z}|\nu)$, so training $g_\phi$
enforces the linear model in~\eqref{eq:linear}.

The encoder and the ratio model are therefore optimized together, from the beginning of
training, with the combined objective
\begin{equation}
  \mathcal{L} = \mathcal{L}_\mathrm{supCL} + \alpha\,\mathcal{L}_\mathrm{cov},
  \label{eq:total}
\end{equation}
where $\alpha > 0$ balances the two terms. Backpropagating through $g_\phi$ into $f_\theta$ encourages the encoder to learn a representation in which the nuisance dependence of the density ratio is well described by the linear nuisance model.

\subsection{Stage 2: Fine-tuning and validating the linear nuisance model with NPLM}
\label{sec:stage2}

\begin{figure*}[!]
    \centering
    \includegraphics[width=\textwidth]
    {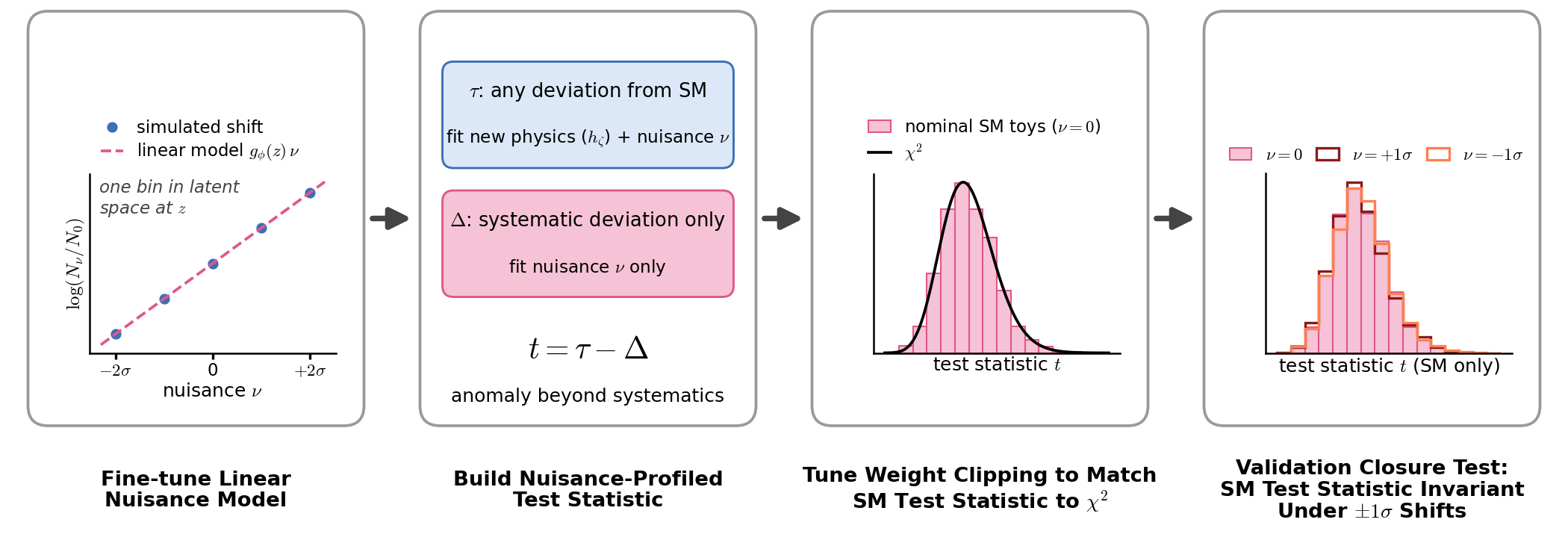}
    \caption{
    \textbf{Treatment of systematic uncertainties in the NPLM framework.}
    From left to right:
    (1) Fine-tune linear model $g_\phi(\mathbf{z})\nu$ to capture the nuisance dependence of $\log(N_\nu/N_0)$, where $N_\nu$ and $N_0$ are shifted and nominal SM counts in latent-space. The nuisance parameter $\nu$ has standard uncertainty $\sigma$.
    (2) Construct $t=\tau-\Delta$: $\tau$ allows both SM  departures, parametrized by the alternative-hypothesis network $h_\zeta$, and nuisance shifts, while $\Delta$ allows nuisance shifts alone.
    (3) Tune the weight clipping of $h_\zeta$ for agreement  between nominal SM pseudo-experiments (toys) and the expected $\chi^2$ distribution.
    (4) Validate closure by checking agreement of the background-only $t$ distributions at $\nu=0,\pm1\sigma$. 
    All distributions are schematic.}
    \label{fig:stage2}
\end{figure*}

Once this joint optimization has converged, we freeze the embedding and validate the linear nuisance model within the NPLM framework, following the strategy outlined in~\cite{dAgnolo:2021aun}. The procedure, summarized in Fig.~\ref{fig:stage2}, consists of four steps:
we fine-tune the linear model, use it to build the profiled test statistic, tune the weight clipping of test network $h_\zeta$ for $\chi^2$ compatibility, and finally validate the modeling via closure checks.

With the encoder frozen, we first fine-tune $g_\phi$ further
to reach the accuracy needed for the statistical test.
The frozen encoder $f_\theta$ and the fine-tuned linear model
$g_\phi$ then integrate directly into the profile-likelihood
NPLM framework described in Section~\ref{sec:nplm}.
The test statistic is the profile-likelihood ratio defined
in Eq.~\eqref{eq:teststat}, evaluated in the latent space
$\mathbf{z}$, with $g_\phi$ modeling the density ratio
$r(\mathbf{z}|\nu)$ given in Eq.~\ref{eq:ratio}.

To calibrate the test, we identify an architecture and
regularization for the test network $h_\zeta$ that render
the test statistic approximately $\chi^2$-distributed under
the null in the absence of uncertainties. We then incorporate
the systematic via the linear model of Eq.~\eqref{eq:linear}
and validate the modeling via closure checks.
We describe the regularization and closure checks below.



\paragraph{Weight clipping for regularization.}
\label{sec:weightclipping}
The flexibility of the test network $h_\zeta$ must be controlled for the test to behave
well. Left unconstrained, the maximum-likelihood fit in Eq.~\eqref{eq:teststat} is
unbounded: $h_\zeta$ can grow without limit to chase the statistical fluctuations of a
finite sample, overfitting the data and driving the null distribution of $t$ into heavy,
non-universal tails that spoil the asymptotic behavior. Following the NPLM
prescription~\cite{DAgnolo:2019vbw, dAgnolo:2021aun}, we regularize the fit by \emph{weight clipping} such that
after every gradient update, the trainable weights of $h_\zeta$ are projected back onto a
bounded box $|w|\le W$, where the weight-clipping parameter $W$ caps the effective capacity
of the network. This bounds the achievable log-density-ratio, and, although the formal
conditions of Wilks' theorem are not satisfied in this setting, we find the null distribution remains close to the asymptotic regime. As a consequence, the function $P(t\,|\,R_{\nu_0})$
approaches a $\chi^2$ whose number of degrees of freedom matches the number of trainable
parameters of $h_\zeta$.

The value of $W$ is a hyperparameter that trades calibration against sensitivity: if $W$ is too small, $h_\zeta$ is too rigid to capture genuine departures and the NPLM test loses power; if $W$ is too large, the network overfits and the null distribution drifts away from the asymptotic $\chi^2$. 
Following~\cite{DAgnolo:2019vbw, dAgnolo:2021aun}, we choose the weight clipping for which the empirical null
distribution---built from an ensemble of central value Reference pseudo-experiments---remains statistically compatible with a $\chi^2$ with number of degrees of freedom matching the number of trainable parameters. 
The value of W selected this way is then used in the computation of likelihood.

 \paragraph{Closure tests for the linear nuisance model}
\label{sec:closure}
The choice of weight clipping and the fine tuning of the linear model $g_\phi$ fully determine the statistical model used in the test. The next step is to check the robustness of the latter to systematic variations of the data. In particular, the test statistics is considered robust to systematic effects if its distribution under the null does not change under mild systematic variations. Operationally, 
\begin{equation}
    \partial_{\nu}P(t|R_\nu)=0 \quad \forall \nu \in [-\sigma, +\sigma].
\end{equation}
To verify this property, 
we build the null distribution $P(t\,|\,R_{\nu_0})$ from an ensemble of pseudo-experiments (\emph{toys}) drawn from the central-value reference. The analogous distribution $P(t\,|\,R_\nu)$ is ontained from toys drawn from a reference with nuisance shifted of $\pm1\sigma$ from the central value. We then quantify their agreement with a two-sample Kolmogorov--Smirnov (KS) test. 
The resulting KS $p$-value measures whether the two \emph{empirical} distributions are statistically compatible. A large value confirms that the profiled nuisance model absorbs the systematic shift, i.e. that $P(t\,|\,R_\nu)$ is effectively independent of $\nu$, as anticipated by the $\tau-\Delta$ cancellation. This construction uses the entire distribution on both sides,
each estimated from a large set of toys.
We refine the linear model and perform this check recursively, until a good agreement is found.

\subsection{Stage 3: Signal injection and anomaly detection}
\label{sec:stage3}
After the test has passed the robustness checks it is ready to be deployed for anomaly detection.
When applied to real data $\cal D$, a single value of the test statistic $t_{\rm obs}$ is observed and thus a single global p-value $p({\cal D})$ is reported:
\begin{equation}
  p = \int_{t_{\rm obs}}^{\infty} P(t\,|\,{\rm null})\,dt.
  \label{eq:pvalue}
\end{equation}

To assess the expected sensitivity to a simulated signal, we apply the test to many pseudo-experiments containing both background and signal. This gives the distribution of the test statistic under the alternative hypothesis. We then estimate the power at a given false-positive rate as the fraction of these pseudo-experiments in which the null hypothesis is rejected. We also report the p-value corresponding to the median test statistic, calculated using Eq. (15), as a measure of typical sensitivity. For probabilities estimated from finite samples of pseudo-experiments, we report Clopper–Pearson confidence intervals.
Throughout this article we will often refer to p-values by their significance expressed as $Z=\Phi^{-1}(1-p)$, with $\Phi$ the standard-normal cumulative distribution function. 

We evaluate this against three choices of null: (a) the empirical distribution $P(t\,|\,R_{\nu_0})$ from central-value toys; (b) the empirical distributions $P(t\,|\,R_\nu)$ of the nuisance-shifted references at $+1\sigma$ and $-1\sigma$; and (c) the analytic $\chi^2$ survival function of the asymptotic distribution discussed above. In all three cases, we used the \emph{median} of the signal distribution, exactly as the significance would be quoted for a single observed dataset. To assess how residual nuisance dependence affects the expected significance, we compare the nominal null distribution with those obtained for nuisance shifts of $\pm1\sigma$. Although the test statistic is designed to be approximately independent of the nuisance parameter, small differences may remain, particularly in the tails of these distributions. At the same median test statistic, a heavier tail gives a larger $p$-value and lower significance. Comparing the nominal and shifted null distributions allows us to check whether using only the nominal reference would overstate the expected sensitivity. Close agreement among (a), (b), and (c) then doubles as a check that $P(t\,|\,R_\nu)$ is effectively $\nu$-independent, so that the single calibration on $R_{\nu_0}$ remains valid across the allowed range of the nuisance. In Section~\ref{sec:experiments} we use the KS construction as a closure test on nuisance-shifted backgrounds and the median-$Z$ construction as our sensitivity measure for injected signals.

\section{Numerical Experiments on LHC Data}
\label{sec:experiments}
In this section, we apply the approach proposed in Section~\ref{sec:method} to a simulated LHC dataset. We evaluate the role of co-training in achieving closure by comparing the test-statistic distributions under nominal and nuisance-shifted background conditions. We also study its impact on signal sensitivity, considering both an idealized setting without nuisance parameters in the NPLM test and a nuisance-aware setting that includes them.

\subsection{Dataset}
We apply the method to simulated CMS Level-1 trigger data~\cite{govorkova2022lhc} available on Zenodo~\cite{thea_aarrestad_2021_5046389, thea_aarrestad_2021_5046446,thea_aarrestad_2021_5055454,thea_aarrestad_2021_5061633, thea_aarrestad_2021_5061688}. Each event is represented by up to 19 reconstructed objects: 4 electrons/photons, 4 muons, 10 jets, and one missing transverse energy (MET) slot. Each object is described by \((p_T,\eta,\phi)\), yielding 57 input features per event, with zero padding applied when fewer objects are present. The background sample consists of \(W\) (\(59.2\%\)), QCD multijet (\(33.8\%\)), \(Z\) (\(6.7\%\)), and \(t\bar{t}\) (\(0.3\%\)) events, while the signal benchmarks are \(\mathrm{LQ}\to b\tau\), \(A\to 4\ell\), \(h^0\to\tau\tau\), and \(h^\pm\to\tau^\pm\nu\).

To model a realistic systematic effect, we introduce a nuisance parameter \(\nu\) that rescales the jet transverse momentum as \(p_T(\nu)=e^\nu p_T(0)\), with nominal value \(\nu_0=0\). We set the standard uncertainty to \(\sigma= 2.5\%\), a typical percent-level size of jet energy scale uncertainties in LHC analyses~\cite{ATLAS_JES}.

After rescaling, we apply the same jet-level cut of $p_T \geq 23\,\mathrm{GeV}$ to the nominal and nuisance-shifted samples, giving them a common selection threshold. This ordering reflects an experimental selection, where trigger or analysis cuts act on the measured jet momenta, which already include any underlying systematic shift. Jets can therefore move across the selection threshold
as $\nu$ varies.

\begin{figure*}[t]
    \setlength{\textfloatsep}{2pt}
    \centering
    \includegraphics[width=\linewidth]{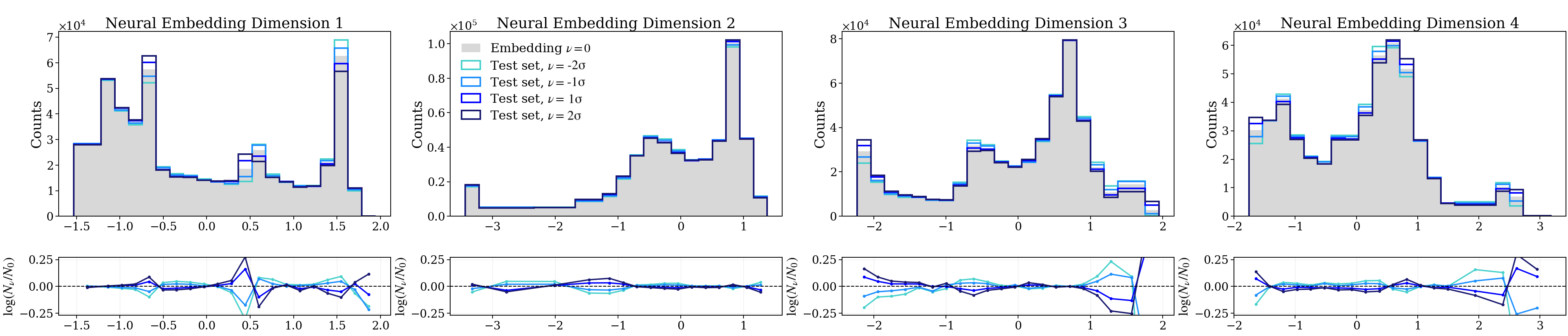}
    \caption{
    \textbf{Unlinearized contrastive embeddings.}
    Marginal distributions of the 4-dimensional learned embedding for the nominal reference sample, $\nu=0$ (gray filled histogram), and nuisance-shifted test samples (colored step histograms). The lower panels show the corresponding binned log-ratios, $\log(N_\nu/N_0)$. The nuisance-shifted distributions do not vary in a linear way across values of $\nu$, making their deformation in the embedding space difficult to capture with the linear nuisance model $g_\phi$.
    }
    \label{fig:embedsb4}
    \vspace{0.5em}
    \includegraphics[width=\linewidth]{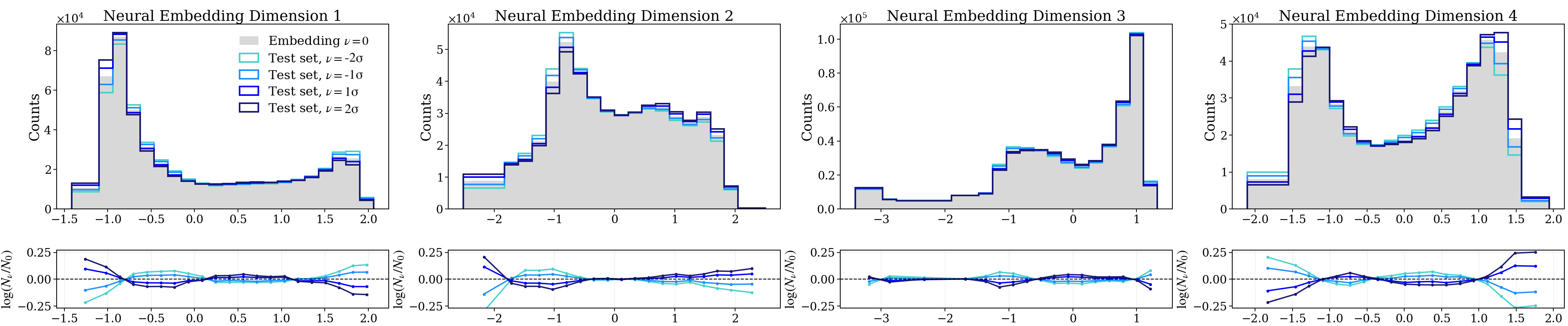}
    \caption{
    \textbf{Linearized contrastive embeddings.}
    Same as Figure~\ref{fig:embedsb4} for a contrastive embedding regularized with the covariance loss $\mathcal{L}_\mathrm{cov}$. The nuisance-shifted samples exhibit a smoother and more coherent dependence on $\nu$. The improved linear covariance makes the nuisance variation easier to model with linear nuisance model $g_\phi$.
    }
    \label{fig:embedslinear}
\end{figure*}

\subsection{Contrastive Embedding Models}

Following Ref.~\cite{Metzger:2025ecl}, we use a Transformer-based encoder $f_\theta$ to map the 57-dimensional input to a 4-dimensional embedding, which gave the strongest AD performance among the dimensions tested. We leave the systematic study of nuisance uncertainty in higher-dimensional embeddings to future work. Nominal SM background events are used for the supervised
contrastive task with physics labels. The initial contrastive embedding is shown in Figure~\ref{fig:embedsb4}. Before co-training with the linear nuisance model, the empirical log-ratios $\log(N_\nu/N_0)$ are not always smoothly ordered or
approximately linear across values of $\nu$, making their nuisance dependence difficult to model with the linear log-density-ratio parametrization of Eq.~\eqref{eq:linear}. 

To impose a more controlled nuisance dependence, we performed the regularized training described in Section~\ref{sec:stage1}. The encoder is trained in a 5-fold co-training setup with the linear model $g_\phi$, using \(2.1\times10^6\) events per fold per $\nu$ dataset, 60 epochs, a batch size of 1024, and a learning rate of \(2\times10^{-6}\).The resulting embedding is shown in Figure~\ref{fig:embedslinear}. After co-training with the covariance loss \(\mathcal{L}_\mathrm{cov}\), the nuisance-shifted marginal distributions vary more smoothly and coherently with $\nu$, making the latent deformation easier to describe with the linear log-density-ratio model $g_\phi(\mathbf{z})\cdot\nu$. Although co-training takes longer than standard contrastive training, the encoder is trained only once and then frozen for the statistical tests. The covariance loss \(\mathcal{L}_\mathrm{cov}\) plateaus during training, while the nuisance model $g_\phi$ benefits from longer fine-tuning.

\begin{figure*}[!]
    \setlength{\textfloatsep}{2pt}
    \centering
    \includegraphics[width=\linewidth]{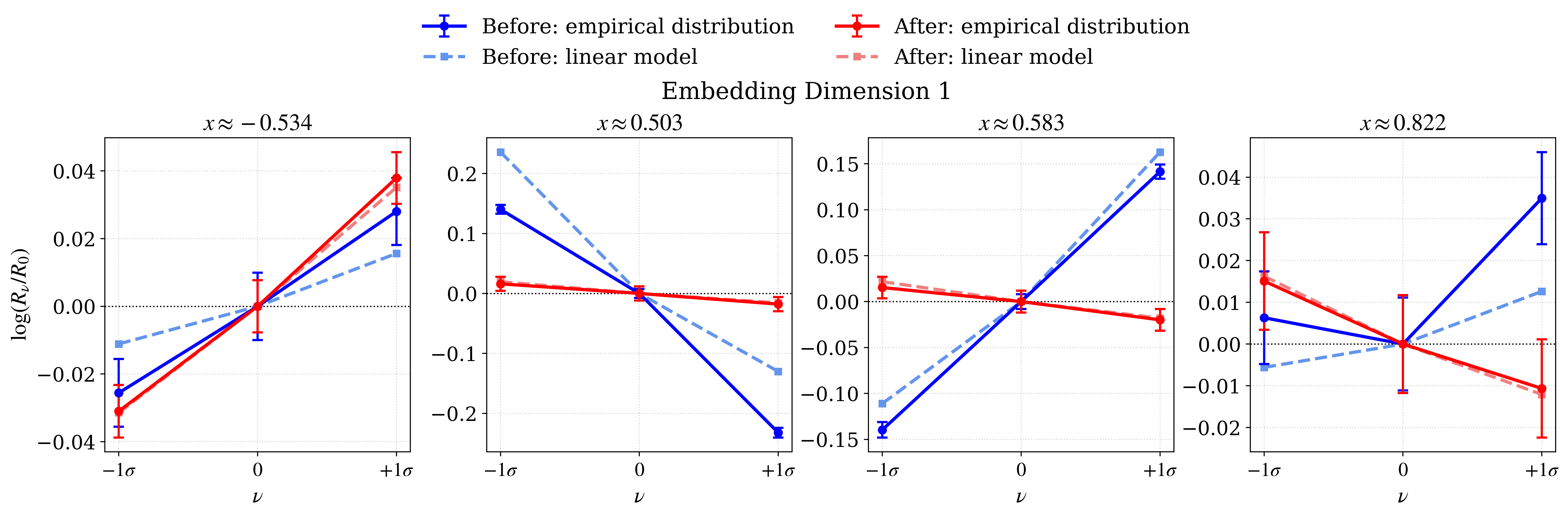}
    \caption{
    \textbf{Empirical and modeled $\log(N_\nu/N_0)$ as a function of the nuisance parameter, before and after embedding linearization.} Each panel shows one bin ($x$) of the first-dimension marginal of the embedding. Solid lines show empirical values at $\nu = -1\sigma$, $0$, and $+\sigma$, with error bars indicating statistical uncertainties; dashed lines show the linear nuisance model predictions. Blue denotes the original embedding and red the embedding trained with the linear regularization. In the absence of linear regularization, the empirical nuisance dependence deviates from the linear model, whereas with regularization the linear assumption is recovered.
    }
    \label{fig:log_sbs}
\end{figure*}
Figure~\ref{fig:log_sbs} directly compares the nuisance dependence of $\log(N_\nu/N_0)$ before and after co-training. Each panel shows a selected bin in the first embedding dimension, with the bin center indicated above the panel. These bins are chosen only to illustrate the effect of linearization and are not representative of every bin in the embedding space. Here, $N_\nu$ is the number of events in that bin for a nuisance-shifted sample, while $N_0$ is the corresponding nominal count. The solid lines with error bars show the empirical values at $\nu=-1\sigma$, $0$, and $+1\sigma$, while the dashed lines show the fits from the linear nuisance model. Results for the original embedding are shown in blue, while those for the linearized embedding are shown in red.

Before co-training, the empirical blue lines can deviate substantially from the corresponding predictions of the linear model, indicating that the nuisance dependence encoded in the original embedding is not sufficiently linear to be accurately described by the linear parameterization of $\log(N_\nu/N_0)$. After co-training, the empirical red lines follow the dashed predictions much more closely across the selected bins, demonstrating a clear improvement in the quality of the linear fit. These results show that jointly training the contrastive embedding and linear nuisance models reshapes the embedding such that its nuisance dependence is more accurately described by $g_\phi(\mathbf{z})\cdot\nu$. In this sense, co-training effectively linearizes the nuisance response of the embedding, making the chosen linear parameterization a substantially better approximation of the observed variations.

\begin{figure*}[t]
    \setlength{\textfloatsep}{2pt}
    \centering
    \includegraphics[width=\linewidth]{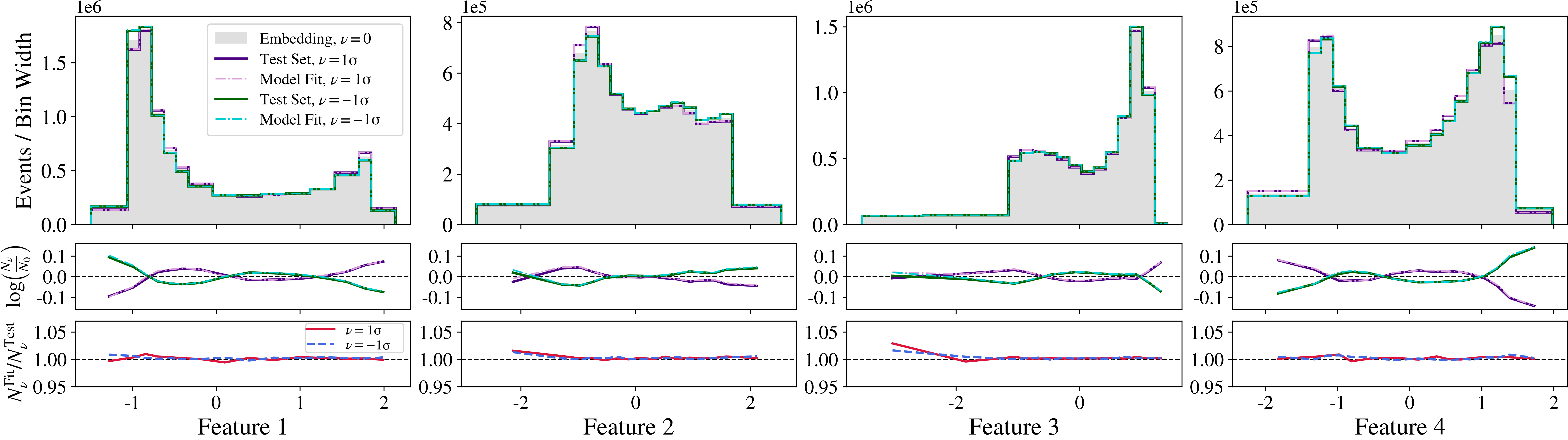}
   \caption{\textbf{Validation of the linear nuisance model in the latent space.} Each column corresponds to one of the four embedding dimensions. Top: distributions for the nominal sample at $\nu = 0$ (gray fill), the nuisance-shifted test samples at $\nu = \pm1\sigma$ (solid lines), and the fitted model predictions (dash-dotted lines). Middle: $\log(N_\nu/N_0)$ for the test samples and the fitted model. Bottom: ratio of model predictions to test distributions, $N_\nu^{\mathrm{fit}}/N_\nu^{\mathrm{test}}$. The close agreement in the upper panels and ratios near unity in the bottom panels show that the linear model accurately captures the nuisance dependence in all four dimensions.}
   \label{fig:models} 
   \end{figure*}
 
\subsection{Linear Model}\label{subsec:linear_model}
We co-train the linear nuisance model $g_\phi$ of Eq.~\eqref{eq:linear} on the learned embeddings. It is linear in the nuisance parameter $\nu$ and, as a function of the embedding $\mathbf{z}$, is implemented as a fully connected network with architecture \((4,512,256,128,64,32,1)\), ReLU activations in the hidden layers, and a linear output. It is trained jointly with the contrastive encoder (Eq.~\eqref{eq:total}) using a learning rate of \(2\times10^{-6}\), \(L_2\) regularization of \(10^{-6}\), and co-training weight \(\alpha=0.1\), then further fine-tuned on \(1.6\times10^6\) events per nuisance parameter. Both nominal and nuisance-shifted samples at \(\pm1\sigma\) and \(\pm2\sigma\) are used for this task. Figure~\ref{fig:models} compares the nuisance-shifted test samples with the predictions of the fitted linear model across the four latent dimensions. The top row shows the nominal distribution at $\nu=0$ in gray, together with the test distributions at $\nu=\pm1\sigma$ (solid lines) and their corresponding model predictions (dash-dotted lines). The middle row shows the same comparison in terms of $\log(N_\nu/N_0)$. In both rows, the test and model samples are visually in close agreement across all four latent dimensions, indicating that the model accurately captures how the latent distributions change with $\nu$. This agreement is further confirmed by the bottom-row ratios of the model prediction to the test distribution, $N_\nu^{\mathrm{fit}}/N_\nu^{\mathrm{test}}$, which remain close to one throughout most regions in the latent space. Overall, the close visual agreement demonstrates that the linear nuisance model accurately captures the nuisance dependence of the embedding, suggesting that co-training has successfully linearized the nuisance dependence of the embedding.

\subsection{Weight clipping regularization}
The alternative-hypothesis network $h_\zeta$ in Eq.~\eqref{eq:teststat}is implemented as a fully connected neural network with architecture \((4,4,4,1)\), corresponding to 45 trainable parameters, with sigmoid activations in the hidden layers and a linear output. Following the calibration procedure of Section~\ref{sec:method}, a weight-clipping regularizer is tuned with 400 pseudo-experiments drawn from the nominal reference sample \(R_{\nu_0}\) so that the empirical null distribution matches the asymptotic \(\chi^2_{45}\) (45 degrees of freedom, one per trainable parameter)~\cite{DAgnolo:2019vbw}. Weight clipping bounds the absolute value of the network's trainable parameters during training, which constrains the smoothness of the learned density ratio and prevents the model from overfitting statistical fluctuations in \(D\); tuning the clipping value is what brings the empirical null into fair agreement with the asymptotic \(\chi^2_{45}\). We train using \(52{,}000\) reference events and \(10{,}000\) background events \(D\) per replica.
We tune the weight clipping for two different implementations: (1) NPLM implemented in the original embedding from~\cite{Metzger:2025ecl} as it is before co-training with the linear nuisance model, and (2) NPLM implemented in the embedding co-trained with the linear nuisance model as explained in Section~\ref{subsec:linear_model}. The optimal weight-clipping is 2.12 in the first case and 1.94 in the second case. Table~\ref{tab:validation_x2} reports the results of the Kolmogorov--Smirnov, Anderson--Darling, Cramér--von Mises, and Pearson $\chi^2$ compatibility tests between the analytical $\chi^2_{45}$ distribution and the empirical \(R_0\) test-statistic distributions obtained with the original and linearized embeddings. Pearson $\chi^2$ tests are performed using binnings corresponding to 10 and 25 degrees of freedom, and each embedding is evaluated at its selected weight-clipping value. All of the resulting \(p\)-values are greater than \(0.5\), indicating good agreement with the analytical distribution and supporting the chosen weight-clipping values.

\begin{table}[t]
\centering
\small
\setlength{\tabcolsep}{8pt}
\renewcommand{\arraystretch}{1.3}

\caption{
\textbf{Compatibility between the empirical nominal test-statistic distributions and the analytical $\chi^2_{45}$ distribution.}
We compare the empirical distribution of the test statistic under $R_0$ with the asymptotic $\chi^2_{45}$, for the linearized and unlinearized embeddings. 
The table reports the $p$-values from various goodness-of-fit tests: the Kolmogorov--Smirnov (KS), Anderson--Darling (AD), Cramér--von Mises (CvM), and Pearson $\chi^2$ ($\chi^2_p$) performed using 10 and 25 degrees of freedom (DoF). 
For both embeddings, the empirical distribution is obtained using 400 replicas and the NPLM test is performed accounting for uncertainties, and using the best weight-clipping value.
}
\label{tab:validation_x2}

\begin{tabular}{c|cc}
\hline
Test
& Unlinearized
& Linearized \\
& ($w_{\mathrm{clip}}=2.12$)
& ($w_{\mathrm{clip}}=1.94$) \\
\hline
KS              & $0.7712$ & $0.8302$ \\
AD                & $0.8124$ & $0.9242$ \\
CvM               & $0.7112$ & $0.9288$ \\
$\chi^2_p$ (10 DoF) & $0.9577$ & $0.9237$ \\
$\chi^2_p$ (25 DoF) & $0.8099$ & $0.7315$ \\
\hline
\end{tabular}
\end{table}

\subsection{NPLM Validations}
We perform the closure test described in Section~\ref{sec:stage2} to verify that nuisance-induced background variations are not misidentified as signal-like excesses. Specifically, we apply the NPLM test to nuisance-shifted SM samples $D\sim R_\nu$, using the same luminosity as in the nominal calibration. Since these samples do not contain injected signal, their test-statistic distributions should remain consistent with the central-value null distribution $P(t\mid R_0)$ once nuisance effects are properly modeled.

\begin{figure*}[tbp]
    \centering
    \begin{subfigure}{0.32\linewidth}
        \centering
        \includegraphics[width=\linewidth]{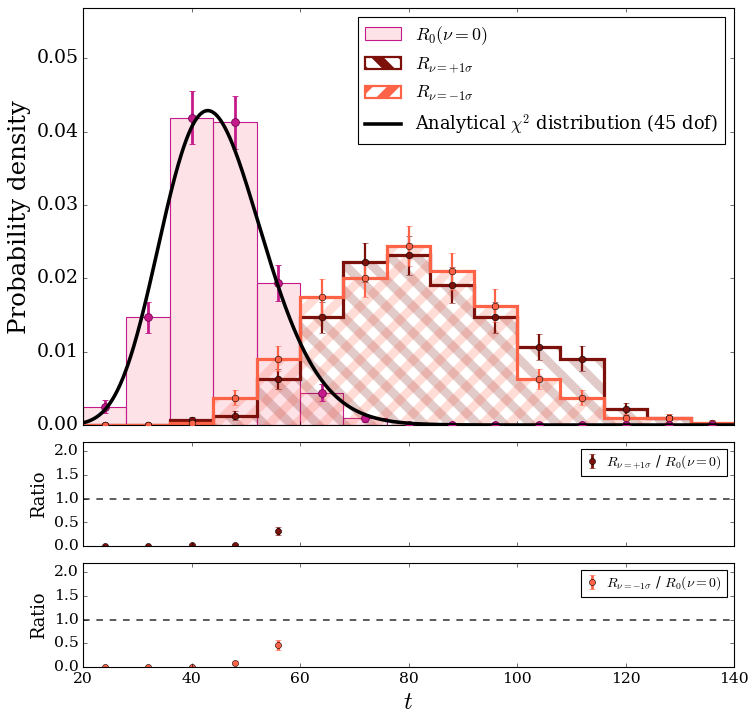}
        \caption{Unlinearized embedding without uncertainties in the NPLM test.}
        \label{fig:null_tstat_closure_a}
    \end{subfigure}
    \hfill
    \begin{subfigure}{0.32\linewidth}
        \centering
        \includegraphics[width=\linewidth]{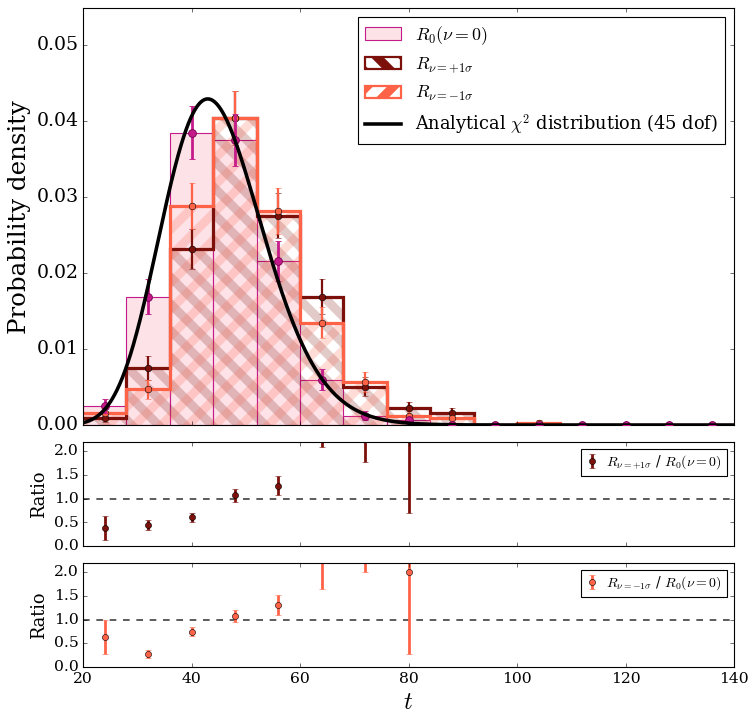}
        \caption{Unlinearized embedding with uncertainties in the NPLM test.}
        \label{fig:null_tstat_closure_b}
    \end{subfigure}
    \hfill
    \begin{subfigure}{0.32\linewidth}
        \centering
        \includegraphics[width=\linewidth]{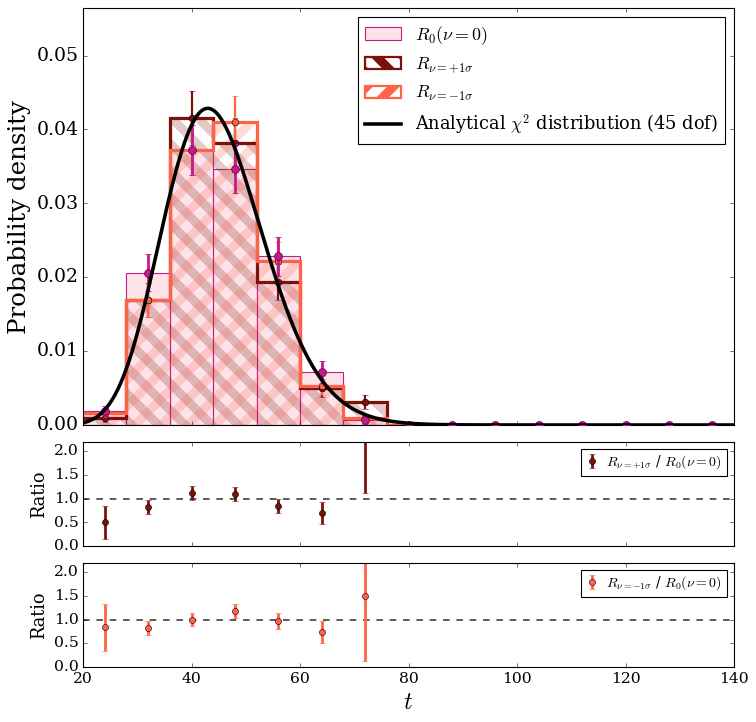}
        \caption{Linearized embedding with uncertainties in the NPLM test (ours).}
        \label{fig:null_tstat_closure_c}
    \end{subfigure}

    \caption{
    \textbf{Background-only closure of the NPLM test statistic in presence of uncertainties.} Each distribution contains 400 background-only replicas. The pink filled histogram shows the empirical central-value distribution $R_0$, while the dark-red and coral step histograms show the nuisance-shifted distributions $R_{\nu=\pm 1\sigma}$; the black curve shows the analytical $\chi^2_{45}$ distribution. The lower panels show the $R_\nu$ to $R_0$ ratios. Panel~\subref{fig:null_tstat_closure_a} shows the NPLM test without nuisance parameters included, for which the nuisance-induced shifts are left unmodeled and appear signal-like. Panel~\subref{fig:null_tstat_closure_b} includes nuisance parameters in the NPLM test but uses the unlinearized embedding, whose nuisance dependence cannot be fully captured by the downstream linear nuisance model, resulting in milder right-shifted $R_{\nu}$ distributions. Panel~\subref{fig:null_tstat_closure_c} is our proposed approach. Using the linearized embedding and the linear nuisance-model in the NPLM test, our approach brings the nominal and nuisance-shifted distributions into agreement.}
    \label{fig:null_tstat_closure}
\end{figure*}

Figure~\ref{fig:null_tstat_closure} compares the background-only test-statistic distributions for $R_0$ and $R_{\nu=\pm 1\sigma}$ under three configurations. In all three panels, the filled red histogram represents the empirical distribution under the central-value reference hypothesis, while the orange and brown step histograms represent the distributions under $R_{\nu=+ 1\sigma}$ and $R_{\nu=- 1\sigma}$, respectively. Panel~\subref{fig:null_tstat_closure_a} shows the NPLM test without treatment of uncertainties. In this case, the test has no way to account for the expected nuisance-induced variations, so the shifted background distributions move away from $R_0$ and can appear signal-like. In panel~\subref{fig:null_tstat_closure_b}, nuisance parameters are included in the NPLM test, but the original unlinearized embedding from Ref.~\cite{Metzger:2025ecl} is used. The shifted distributions still do not reach closure because the nuisance dependence is encoded nonlinearly in the learned representation and cannot be fully absorbed by adding only a downstream linear nuisance model. This motivates linearizing the embedding itself rather than relying on the NPLM nuisance fit alone. Finally, panel~\subref{fig:null_tstat_closure_c} shows the result of our method, co-training the embedding with the linear nuisance model. The nominal and nuisance-shifted distributions are now in good agreement, demonstrating that the linearized embedding, together with the nuisance-aware NPLM test, successfully prevents the systematic variations from causing false positives. 

We quantify the agreement in the final linearized configuration shown in panel~\subref{fig:null_tstat_closure_c} using 400 background-only replicas for each nuisance value. Table~\ref{tab:validation_A} presents two complementary checks. First, we assess closure by comparing the nuisance-shifted test-statistic distributions, $R_{\nu=\pm1\sigma}$, with
the nominal reference $R_0$ using the Kolmogorov--Smirnov (KS), Anderson-Darling (AD), and Cramer-von-Mises (CvM) tests, and binned Pearson $\chi^2$ compatibility tests with 10 and 25 degrees of freedom. All closure-test $p$-values exceed 0.5, indicating that the shifted distributions are consistent with the nominal reference. 
Second, we compare the same nuisance-shifted distributions with the analytical $\chi^2_{45}$ distribution using the KS, Anderson--Darling (AD), Cramér--von Mises (CvM), and Pearson $\chi^2$ tests. These $p$-values range from 0.24 to 0.86 and show no significant departure from the analytical reference. Together, these results support the interpretation that the linear model $g_\phi$, combined with the co-trained embedding, describes the nuisance-induced deformation sufficiently well to preserve the background-only test-statistic distribution. This closure is essential for latent-space anomaly detection, where an unmodeled systematic shift could otherwise produce an artificially significant excess.

To assess the role of co-training in achieving this closure, Table~\ref{tab:validation_unlinearized} reports the same tests for the unlinearized embedding with nuisance parameters included in the NPLM test, corresponding to panel~\subref{fig:null_tstat_closure_b}. Although the nominal distribution agrees with the analytical
$\chi^2_{45}$ reference, both nuisance-shifted distributions are inconsistent with $R_0$ and with the analytical reference, with small $p$-values across all tests. Agreement at the nominal nuisance value therefore does not ensure closure under nuisance shifts. Together with the linearized results, this comparison shows that the downstream linear nuisance model alone is insufficient
to describe the shifts in the original embedding, while co-training enables the same model to achieve closure.

\begin{table}[t]
\centering
\small
\setlength{\tabcolsep}{5pt}
\renewcommand{\arraystretch}{1.3}

\caption{
\textbf{Validation of the NPLM test in the linearized embedding.}
We compare the nuisance-shifted background-only NPLM test-statistic distributions $R_{\nu=\pm1\sigma}$ with two references: the empirical nominal distribution $R_0$ (left block) and the analytical $\chi^2_{45}$ distribution (right block).
The table reports $p$-values from various goodness-of-fit tests: the Kolmogorov--Smirnov (KS), Anderson--Darling (AD), Cramér--von Mises (CvM), and Pearson $\chi^2$ ($\chi^2_p$) performed using 10 and 25 degrees of freedom (DoF).
Each empirical distribution is obtained using 400 replicas.}
\label{tab:validation_A}

\begin{tabular}{c | cc | cc}
\hline
Test
 & \multicolumn{2}{c|}{\textit{empirical $R_0$}}
 & \multicolumn{2}{c}{\textit{analytical $\chi^2_{45}$}} \\
& $\nu=+1\sigma$ & $\nu=-1\sigma$ & $\nu=+1\sigma$ & $\nu=-1\sigma$ \\
\hline
KS
    & $0.7583$ & $0.9416$ & $0.4514$ & $0.4480$ \\
AD
    & $0.6526$ & $0.8287$ & $0.2402$ & $0.4043$ \\
CvM
    & $0.7824$ & $0.8886$ & $0.3917$ & $0.5069$ \\
$\chi^2_p$ (10 DoF)
    & $0.7833$ & $0.8229$ & $0.6727$ & $0.4676$ \\
$\chi^2_p$ (25 DoF)
    & $0.8122$ & $0.7966$ & $0.8605$ & $0.4352$ \\
\hline
\end{tabular}
\end{table}

\begin{table}[t]
\centering
\small
\setlength{\tabcolsep}{5pt}
\renewcommand{\arraystretch}{1.3}

\caption{
\textbf{Validation of the NPLM test in the unlinearized embedding.}
With nuisance parameters included in the NPLM test, we compare the nuisance-shifted background-only test-statistic distributions $R_{\nu=\pm1\sigma}$ with the empirical nominal distribution
$R_0$ and the analytical $\chi^2_{45}$ distribution. Each empirical distribution contains 400 replicas. The small $p$-values indicate a lack of closure and disagreement with the analytical reference. AD values are reported at the vailable precision.}

\label{tab:validation_unlinearized}

\resizebox{\columnwidth}{!}{%
\begin{tabular}{c | cc | cc}
\hline
Test
 & \multicolumn{2}{c|}{\textit{empirical $R_0$}}
 & \multicolumn{2}{c}{\textit{analytical $\chi^2_{45}$}} \\
& $\nu=+1\sigma$ & $\nu=-1\sigma$
& $\nu=+1\sigma$ & $\nu=-1\sigma$ \\
\hline
KS
    & $5.806\times10^{-14}$ & $1.371\times10^{-7}$
    & $2.555\times10^{-28}$ & $2.870\times10^{-17}$ \\
AD
    & $1.000\times10^{-4}$ & $1.000\times10^{-4}$ & $1.000\times10^{-4}$ & $1.000\times10^{-4}$ \\
CvM
    & $<10^{-6}$ & $<10^{-6}$ & $<10^{-6}$ & $<10^{-6}$ \\
$\chi^2_p$ (10 DoF)
    & $1.227\times10^{-10}$ & $3.374\times10^{-8}$
    & $1.054\times10^{-36}$ & $1.532\times10^{-21}$ \\
$\chi^2_p$ (25 DoF)
    & $7.566\times10^{-10}$ & $3.704\times10^{-6}$
    & $3.393\times10^{-32}$ & $1.770\times10^{-20}$ \\
\hline
\end{tabular}%
}
\end{table}
 
\subsection{Signal Sensitivity}

\begin{figure*}[]
\centering
\begin{subfigure}[t]{0.32\linewidth}
        \centering
        \includegraphics[width=\linewidth]
        {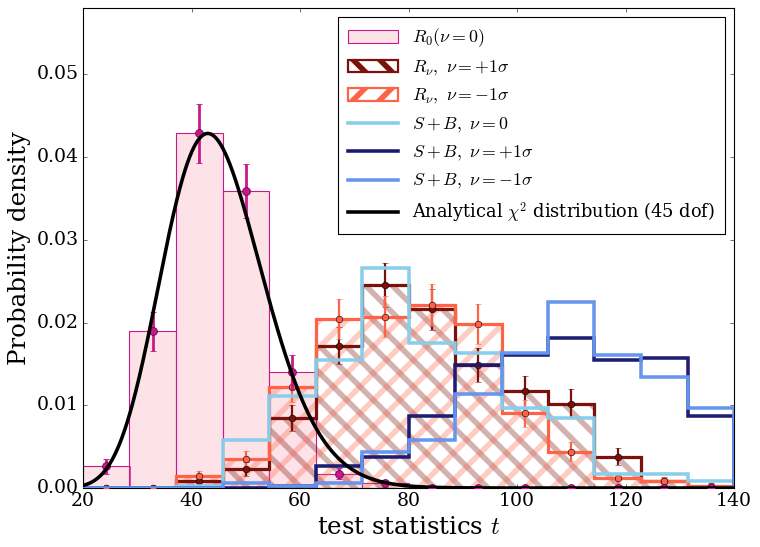}
        \caption{Unlinearized embedding without uncertainties in the NPLM test.}
        \label{fig:signal_tstat_a}
    \end{subfigure}
    \hfill
    \begin{subfigure}[t]{0.32\linewidth}
        \centering
        \includegraphics[width=\linewidth]
        {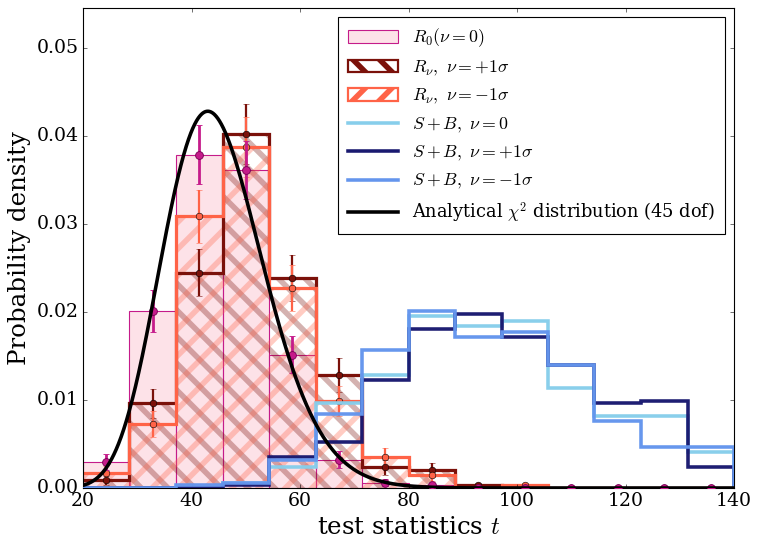}
        \caption{Unlinearized embedding with uncertainties in the NPLM test.}
        \label{fig:signal_tstat_b}
    \end{subfigure}
    \hfill
\begin{subfigure}[t]{0.32\linewidth}
        \centering
        \includegraphics[width=\linewidth]
        {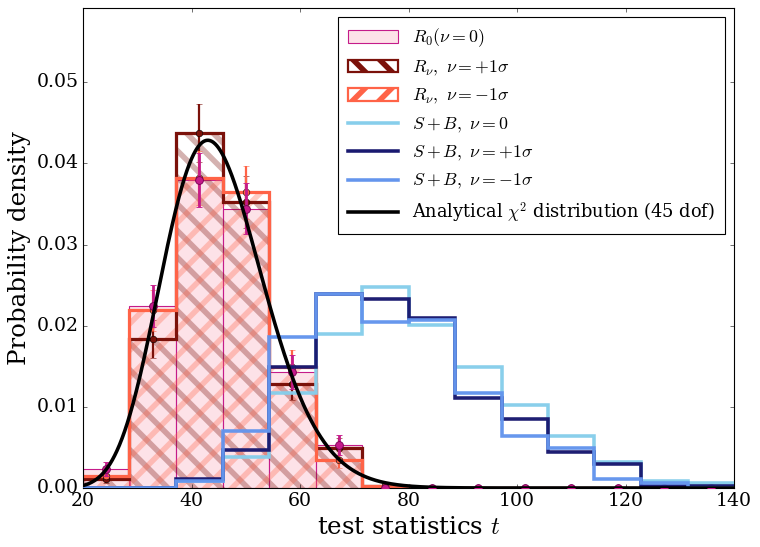}
        \caption{Linearized embedding with uncertainties in the NPLM test (ours).}
        \label{fig:signal_tstat_c}
    \end{subfigure}

\caption{
\textbf{Sensitivity to \(1\%\)
\(H^\pm\to\tau^\pm\nu\) signal injection in presence of uncertainties.}
Each distribution is estimated from 400 replicas. The pink filled histogram shows the nominal
background-only distribution \(R_0\), while the dark-red and coral hatched histograms show the nuisance-shifted background distributions
\(R_{\nu=+ 1\sigma}\) and \(R_{\nu=- 1\sigma}\), respectively. The light-blue, dark-blue, and medium-blue step histograms show the corresponding
signal-plus-background distributions at \(\nu=0\), \(+ 1\sigma\), and
\(- 1\sigma\). The black curve shows the analytical $\chi^2$ distribution with 45 degrees of freedom. Panel~(a) omits nuisance parameters from the NPLM test;
panel~(b) includes the nuisance parameter in the NPLM test, using the original, unlinearized embedding; and
panel~(c) includes the nuisance parameter in the NPLM test, using the co-trained, linearized embedding. The
markers and error bars show the binwise estimates of the background-only densities and their statistical uncertainties.}
\label{fig:signal_hists}
\end{figure*}

We then assess signal sensitivity by injecting a small signal component into the SM background sample. We consider 0.5\% and 1\% signal injections, corresponding in our luminosity settings to 50 and 100 injected signal events, respectively. For each injected sample, we compute the NPLM test statistic and convert it into an empirical $z$-score relative to the chosen background-only calibration distribution, estimated from 400 replicas. Figure~\ref{fig:signal_hists} compares the test-statistic distributions obtained for a $1\%$ $H^\pm\to\tau^\pm\nu$ injection under the three nuisance-treatment configurations considered in this work. The first configuration, Figure~\ref{fig:signal_tstat_a}, omits nuisance parameters from the NPLM test and shows how the test responds when systematic variations are not explicitly accounted for. In the second configuration, shown in Figure~\ref{fig:signal_tstat_b}, the nuisance parameters are included in the test using the original, unlinearized embedding. Figure~\ref{fig:signal_tstat_c} combines the nuisance-aware NPLM test with the co-trained, linearized embedding and therefore corresponds to the full procedure proposed in this work.

The distributions in Figure~\ref{fig:signal_hists} show that, when nuisance parameters are omitted from the NPLM test, the test statistic distribution under nuisance-shifted background-only (orange and brown) and under signal-injection (shades of blue) both shift toward larger values and cannot be reliably distinguished. Including the nuisance parameters with the original, unlinearized embedding also does not fully resolve this problem, as shown in Figure~\ref{fig:signal_tstat_b}, where the shifted background distributions remain separated from $R_0$ and alter the calculation of p-values. On the other hand, after co-training, the background-only distributions achieve closure, while the signal-injected distributions remain right-shifted toward larger $t$ for all tested nuisance values. The linearized embedding therefore allows nuisance-induced variations to be calibrated retaining sensitivity to the signal injection across the tested values of $\nu$.

The detection results of our method on experiments with signal-injection across the four signal benchmarks are summarized in Table~\ref{tab:validation_C}. The table reports the median observed $z$-scores obtained by calibrating each test sample against the empirical background-only distribution $R_{\nu=+1\sigma}$. We use the same reference for the unlinearized results in Table~\ref{unlinearized_with_nuisance} to allow a direct comparison. These results show that the NPLM test retains sensitivity to small injected signals across the tested nuisance values. To further show the stability of our results, we report in Appendix~\ref{app:closure} the same study using the nominal reference $R_0$, the nuisance-shifted reference $R_{\nu=-1\sigma}$, and the asymptotic $\chi^2_{45}$ distribution for calibration.

\begin{table*}[!th]
\centering
\small
\setlength{\tabcolsep}{5.5pt}
\renewcommand{\arraystretch}{1.6}
\caption{
\textbf{Sensitivity to signal benchmarks using the linearized embedding and including the uncertainties in the NPLM test.}
Median observed $z$-scores for each injected signal benchmark and test-sample nuisance value, calibrated against the empirical background-only distribution obtained with $+1\sigma$ ($R_{\nu=+1\sigma}$). Each signal benchmark is evaluated at $0.5\%$ and $1\%$ signal injection, shown in the left and right columns, respectively. For each benchmark and injection fraction, we generate 400 null toys and 400 signal-injected toys. Uncertainties denote 68\% Clopper--Pearson confidence
intervals. Values reported as $\geq 2.81$ saturate the finite-sample calibration limit.}
\label{tab:validation_C}

\vspace{0.3em}

\begin{tabular}{c|cc|cc|cc|cc}
\hline
Test sample
& \multicolumn{2}{c|}{$LQ\to bt$}
& \multicolumn{2}{c|}{$A\to 4\ell$}
& \multicolumn{2}{c|}{$h\to \tau\tau$}
& \multicolumn{2}{c}{$H^\pm\to \tau^\pm\nu$} \\
\cline{2-9}
\footnotesize{(calibrated by $R_{\nu=+1\sigma}$)}
& $0.5\%$ & $1\%$
& $0.5\%$ & $1\%$
& $0.5\%$ & $1\%$
& $0.5\%$ & $1\%$ \\
\hline

$\nu=0$
& $0.29^{+0.07}_{-0.06}$ & $1.08^{+0.09}_{-0.07}$
& $0.82^{+0.08}_{-0.07}$ & $\geq 2.81$
& $0.11^{+0.07}_{-0.06}$ & $0.65^{+0.08}_{-0.07}$
& $1.01^{+0.09}_{-0.07}$ & $\geq 2.81$ \\

$\nu=+1\sigma$
& $0.28^{+0.07}_{-0.06}$ & $0.98^{+0.09}_{-0.07}$
& $0.91^{+0.08}_{-0.07}$ & $\geq 2.81$
& $0.17^{+0.07}_{-0.06}$ & $0.61^{+0.08}_{-0.07}$
& $0.99^{+0.09}_{-0.07}$ & $\geq 2.81$ \\

$\nu=-1\sigma$
& $0.24^{+0.07}_{-0.06}$ & $0.93^{+0.09}_{-0.07}$
& $0.86^{+0.08}_{-0.07}$ & $\geq 2.81$
& $0.17^{+0.07}_{-0.06}$ & $0.61^{+0.07}_{-0.06}$
& $0.85^{+0.08}_{-0.07}$ & $\geq 2.81$ \\
\hline
\end{tabular}
\end{table*}


\begin{table*}[!th]
\centering
\small
\setlength{\tabcolsep}{5.5pt}
\renewcommand{\arraystretch}{1.6}
\caption{
\textbf{Signal-injection $z$-scores using the original, unlinearized embeddings
with nuisance parameters included in the NPLM test.}
Median observed $z$-scores are reported for each injected signal benchmark and test-sample nuisance value. Each signal benchmark is evaluated at $0.5\%$ and $1\%$ signal injection, shown in the left and right columns, respectively. For each benchmark and injection fraction, we generate 400 null toys and 400 signal-injected toys. The $z$-scores are calibrated against the empirical background-only distribution $R_{\nu=+1\sigma}$. This distribution has the longest high-$t$ tail among the tested nuisance values and therefore provides the most conservative calibration when the unlinearized embedding does not achieve closure between $R_0$ and $R_\nu$. Negative $z$-scores occur when the median test statistic of a signal-injected distribution lies below the median of this conservative reference, corresponding to a one-sided empirical $p$-value greater than $0.5$. Uncertainties denote 68\% Clopper--Pearson confidence intervals. Values reported as $\geq 2.81$ saturate the finite-sample calibration limit.
}

\label{unlinearized_with_nuisance}
\vspace{0.3em}
\begin{tabular}{c|cc|cc|cc|cc}
\hline
Test sample
& \multicolumn{2}{c|}{$LQ\to bt$}
& \multicolumn{2}{c|}{$A\to 4\ell$}
& \multicolumn{2}{c|}{$h\to \tau\tau$}
& \multicolumn{2}{c}{$H^\pm\to \tau^\pm\nu$} \\
\cline{2-9}
\footnotesize{(calibrated by $R_{\nu=+1\sigma}$)}
& $0.5\%$ & $1\%$
& $0.5\%$ & $1\%$
& $0.5\%$ & $1\%$
& $0.5\%$ & $1\%$ \\
\hline
$\nu=0$
& $-0.58^{+0.07}_{-0.07}$ & $0.97^{+0.09}_{-0.07}$
& $0.39^{+0.07}_{-0.06}$ & $2.24^{+0.32}_{-0.14}$
& $-0.12^{+0.07}_{-0.06}$ & $0.39^{+0.07}_{-0.06}$
& $0.72^{+0.08}_{-0.07}$ & $\geq 2.81$ \\

$\nu=+1\sigma$
& $0.37^{+0.07}_{-0.06}$ & $1.46^{+0.12}_{-0.09}$
& $0.68^{+0.08}_{-0.07}$ & $2.24^{+0.32}_{-0.14}$
& $0.32^{+0.07}_{-0.06}$ & $0.96^{+0.09}_{-0.07}$
& $1.14^{+0.10}_{-0.08}$ & $\geq 2.81$ \\

$\nu=-1\sigma$
& $0.35^{+0.07}_{-0.06}$ & $1.19^{+0.10}_{-0.08}$
& $0.67^{+0.08}_{-0.07}$ & $2.17^{+0.28}_{-0.13}$
& $0.10^{+0.07}_{-0.06}$ & $0.76^{+0.08}_{-0.07}$
& $0.96^{+0.09}_{-0.07}$ & $\geq 2.81$ \\

\hline
\end{tabular}
\end{table*}

Table~\ref{unlinearized_with_nuisance} instead evaluates the original embedding after nuisance parameters are incorporated into the NPLM test. Because the unlinearized embedding does not achieve closure between $R_0$ and $R_\nu$, the nominal distribution $R_0$ cannot be treated as a nuisance-independent null reference. 
Among the tested reference distributions, $R_{\nu=+ 1\sigma}$ has the longest high-$t$ tail and yields slightly more conservative significances. We therefore calibrate these results against $R_{\nu=+ 1\sigma}$ as the conservative reference where closure is not achieved. 
Even under this conservative calibration, the significance of the same signal injection remains strongly dependent on the nuisance value of the test sample. Although some variation is expected from finite-sample fluctuations and further reduced sensitivity can be caused by nuisance coupling with the injected signal, the variations observed across nuisance values within the same signal-injection column in Table~\ref{unlinearized_with_nuisance} indicate that the reported sensitivity is strongly driven by the lack of background closure. In particular, the sensitivity is systematically larger when the nuisance is not in the central value, suggesting that the observed z-score is inflated by the residual mismodeling of the uncertainties. 
The instability is further reflected in the negative $z$-scores, which
occur when the median test statistic of a signal-injected distribution lies below the median of the conservative $R_{\nu=+ 1\sigma}$ reference distribution, corresponding to a one-sided empirical $p$-value greater than $0.5$. In these cases, the nuisance-induced displacement of the unlinearized background distribution is larger than the displacement produced by the injected signal. The negative values therefore do not indicate a ``negative signal,'' but instead demonstrate that a post-hoc correction for the absence of closure causes a systematic shift that can obscure or overwhelm the signal response. Overall, this nuisance dependence makes the resulting significances unreliable. The same underlying signal can appear more significant, less significant, or even shifted below the reference median, causing loss in sensitivity and even chances of false claims.

\begin{figure}[!]
    \centering
    \includegraphics[width=\linewidth]
{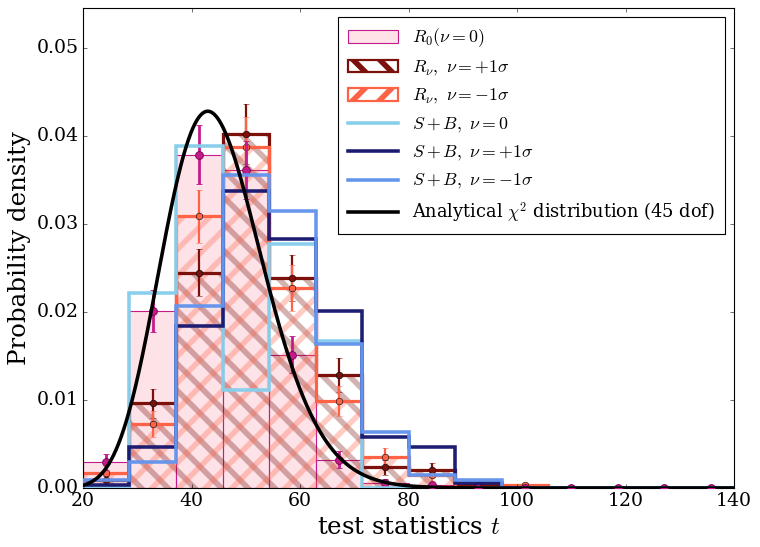}
    \caption{
\textbf{Nuisance-induced shifts can overwhelm the signal response.}
Test-statistic distributions in the unlinearized embedding for a $0.5\%$ $LQ\to bt$ injection, with nuisance parameters included in the NPLM test. Pink and red histograms show the nominal and nuisance-shifted SM backgrounds; blue histograms show signal-injected samples. The black curve shows the analytical $\chi^2_{45}$ distribution.The nominal signal-injected median lies below that of $R_{\nu=+1\sigma}$, yielding a negative $Z$-score when calibrated against this reference.}
    \label{fig:lq_unlinearized_overlap}
\end{figure}

\begin{figure*}[t]
    \centering
    \captionsetup[subfigure]{skip=1pt}

    {\small\textbf{Unlinearized embedding}\par}
    \vspace{0.2em}

    \begin{subfigure}[t]{0.247\textwidth}
        \centering
        \includegraphics[width=\linewidth,trim=6 6 6 22.5,clip]
        {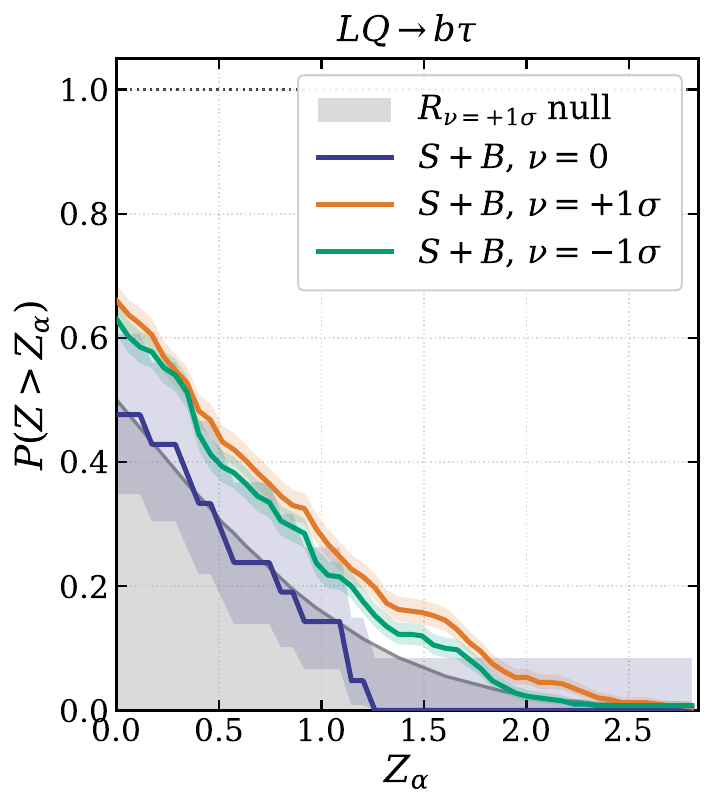}
        \caption{$LQ\to bt$}
    \end{subfigure}\hfill
    \begin{subfigure}[t]{0.247\textwidth}
        \centering
        \includegraphics[width=\linewidth,trim=6 6 6 22.5,clip]
        {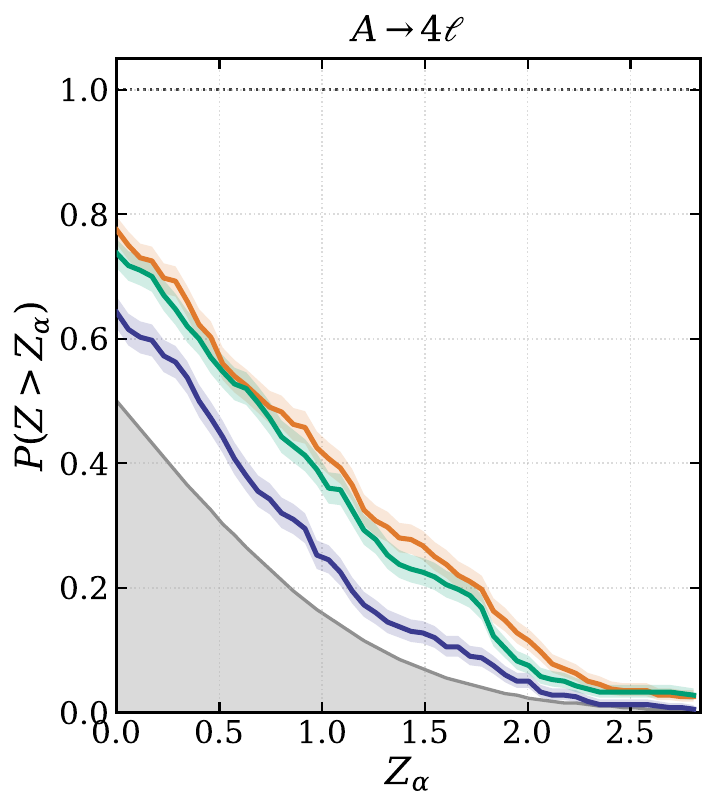}
        \caption{$A\to 4\ell$}
    \end{subfigure}\hfill
    \begin{subfigure}[t]{0.247\textwidth}
        \centering
        \includegraphics[width=\linewidth,trim=6 6 6 22.5,clip]
        {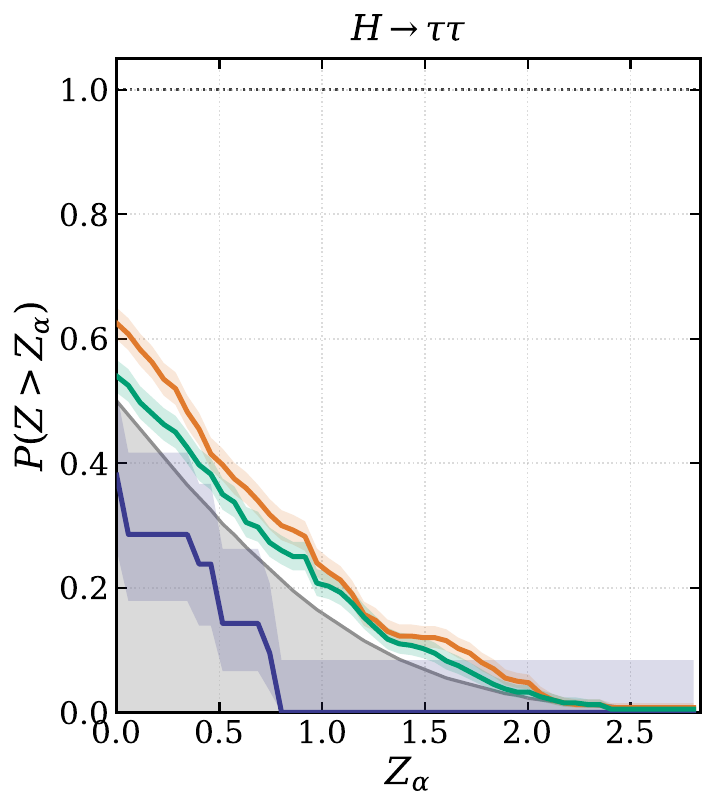}
        \caption{$h\to \tau\tau$}
    \end{subfigure}\hfill
    \begin{subfigure}[t]{0.247\textwidth}
        \centering
        \includegraphics[width=\linewidth,trim=6 6 6 22.5,clip]
        {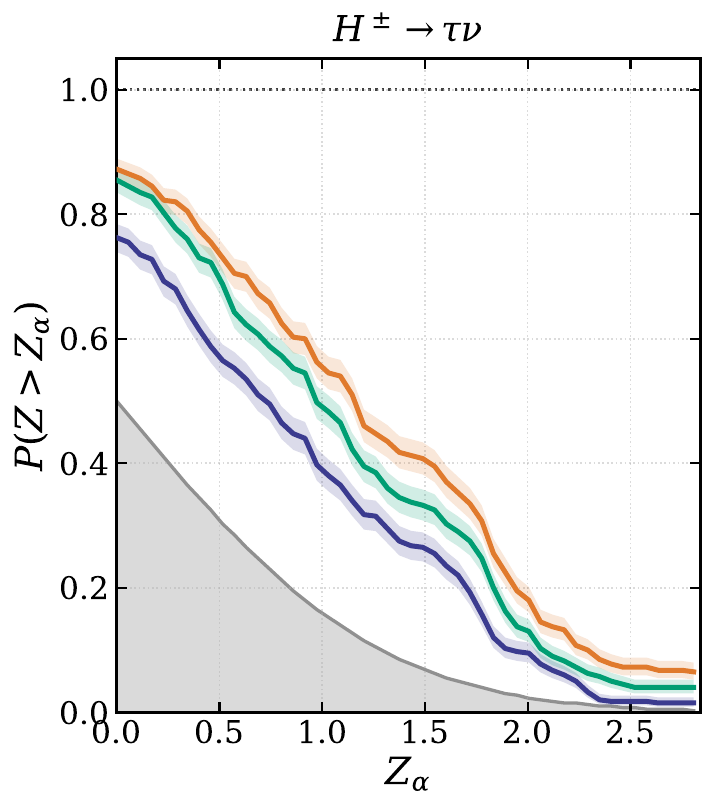}
        \caption{$H^\pm\to \tau^\pm\nu$}
    \end{subfigure}

    \par\vspace{0.6em}

    {\small\textbf{Linearized embedding}\par}
    \vspace{0.2em}

    \begin{subfigure}[t]{0.247\textwidth}
        \centering
        \includegraphics[width=\linewidth,trim=6 6 6 22.5,clip]
        {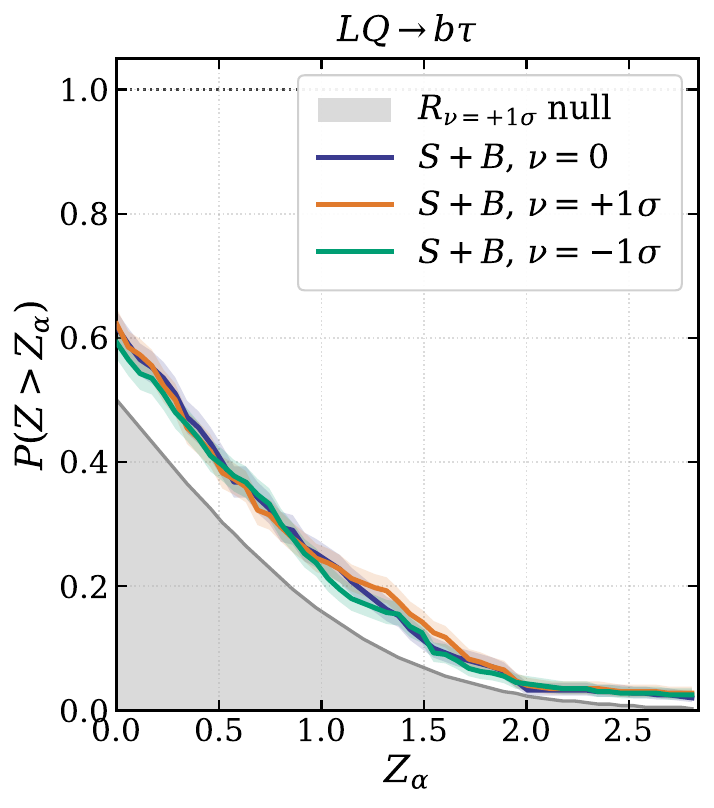}
        \caption{$LQ\to bt$}
    \end{subfigure}\hfill
    \begin{subfigure}[t]{0.247\textwidth}
        \centering
        \includegraphics[width=\linewidth,trim=6 6 6 22.5,clip]
        {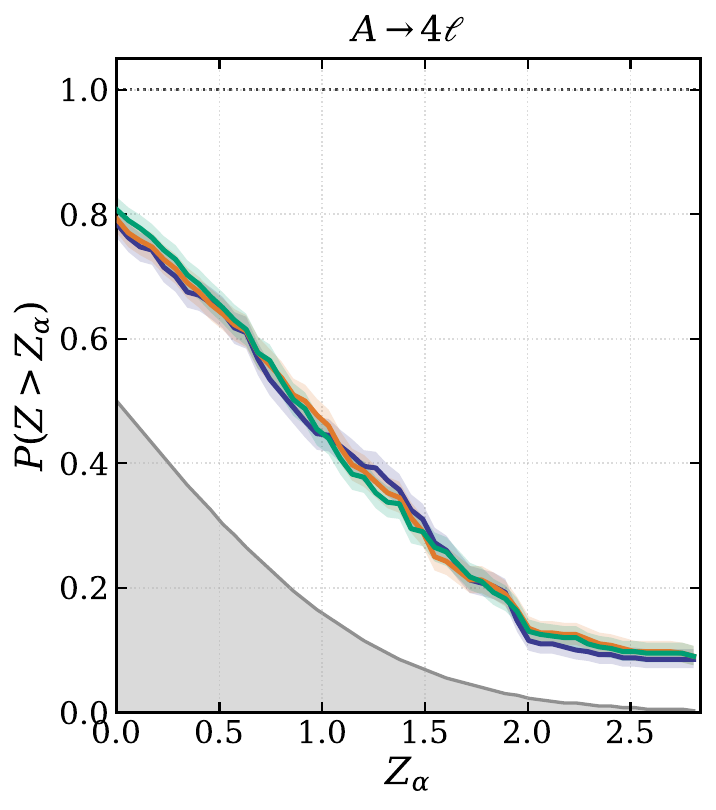}
        \caption{$A\to 4\ell$}
    \end{subfigure}\hfill
    \begin{subfigure}[t]{0.247\textwidth}
        \centering
        \includegraphics[width=\linewidth,trim=6 6 6 22.5,clip]
        {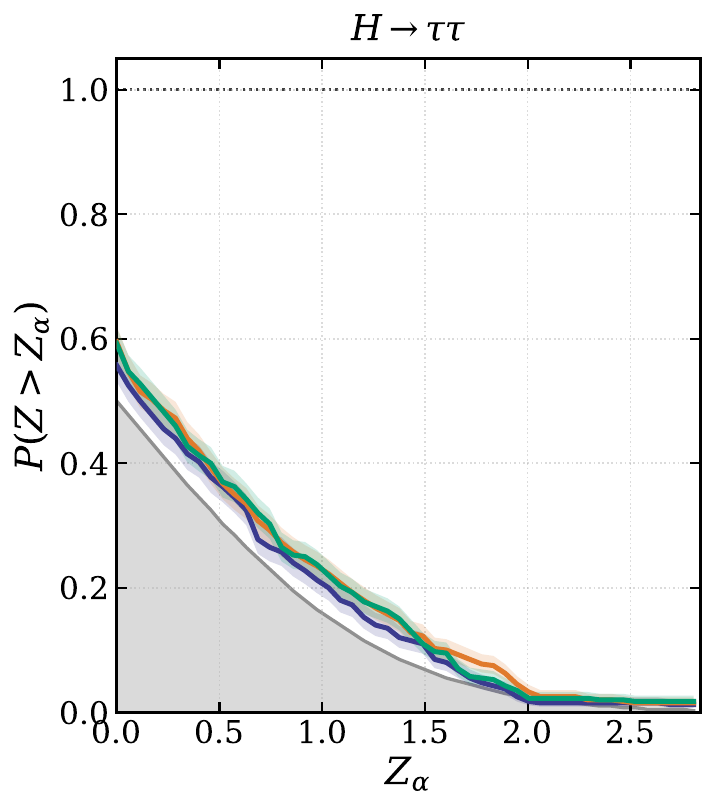}
        \caption{$h\to \tau\tau$}
    \end{subfigure}\hfill
    \begin{subfigure}[t]{0.247\textwidth}
        \centering
        \includegraphics[width=\linewidth,trim=6 6 6 22.5,clip]
        {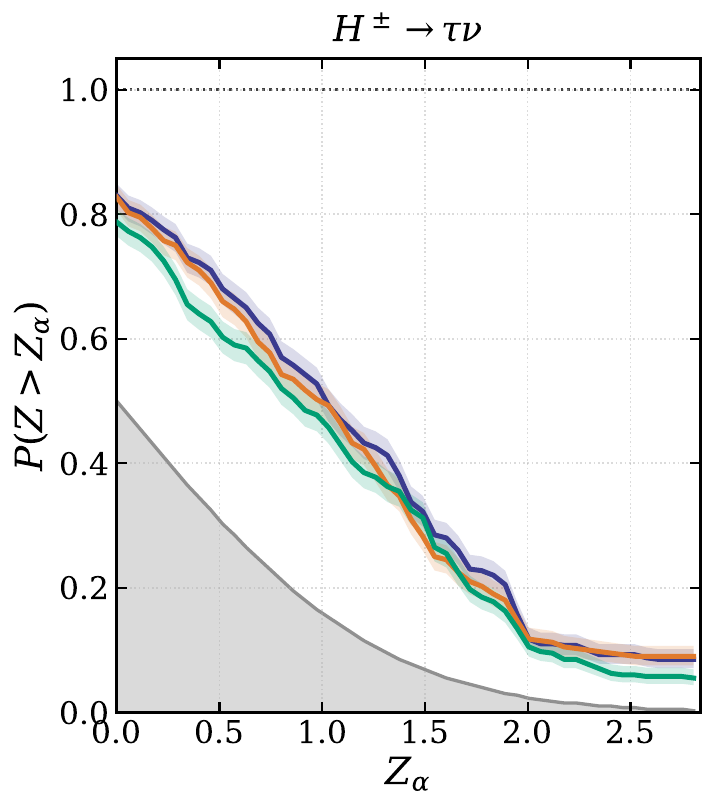}
        \caption{$H^\pm\to \tau^\pm\nu$}
    \end{subfigure}

    \caption{
    \textbf{Power curves for $0.5\%$ signal injection using the unlinearized and linearized embeddings.} The top row shows the unlinearized results, and the bottom row shows the linearized results, both empirically calibrated against $R_{\nu=+1\sigma}$. Each panel is evaluated using 400 null and 400 signal-injected toys. From left to right, the signal benchmarks are $\mathrm{LQ}\to bt$, $A\to 4\ell$, $h\to\tau\tau$, and $H^\pm\to\tau^\pm\nu$. The colored curves correspond to signal-injected samples evaluated at $\nu=0$ and $\nu=\pm1\sigma$. The gray curve corresponds to the SM-only reference sample $R_{\nu=+1\sigma}$, calibrated against its own empirical test-statistic distribution. For the unlinearized embedding, $R_{\nu=+1\sigma}$ has the largest high-$t$ tail among the tested nuisance-shifted SM samples, so using it as the calibration reference gives a more conservative estimate of the expected discovery power. We use the same reference for the linearized results to allow a direct comparison.}
    \label{fig:power_curves_emp_n50}
\end{figure*}

Figure~\ref{fig:lq_unlinearized_overlap} illustrates this effect for a $0.5\%$ injection of the $LQ\to bt$ signal. The background-only distribution at $\nu=+1\sigma$ has a larger median test statistic than the signal-injected
distribution at $\nu=0$. Calibrating the latter against $R_{\nu=+1\sigma}$ therefore gives a median $p$-value above $0.5$ and a negative $Z$-score, despite the presence of signal. This example shows how, without co-training, residual nuisance effects can exceed the response to an injected signal,limiting the test's ability to distinguish new physics from nuisance-induced shifts.

Conversely, the detection performed in the linearized embedding
(Table~\ref{tab:linearized}) yields $z$-scores that are substantially more stable across the tested nuisance values: within each benchmark and injection fraction, the spread across $\nu$ is at most $0.16$, compared with up to $0.95$ for the unlinearized embedding. At the nominal value $\nu=0$, the linearized embedding also yields equal or higher significance for every benchmark. At $\nu=\pm1\sigma$ the unlinearized embedding can return larger $z$-scores, but, as discussed above, these are inflated by residual nuisance mismodeling rather
than reflecting genuine sensitivity. The closure between $R_0$ and $R_\nu$ provides a consistent background-only calibration, ensuring that the inferred significances primarily reflect the injected signals rather than unknown nuisance variations. The linearized embedding therefore provides both improved sensitivity and reliably calibrated significances, making it the more robust and practically usable representation for an uncertainty-aware anomaly search.

\begin{figure*}[t]
    \centering
    \captionsetup[subfigure]{skip=1pt}

    {\small\textbf{Unlinearized embedding}\par}
    \vspace{0.2em}

    \begin{subfigure}[t]{0.247\textwidth}
        \centering
        \includegraphics[width=\linewidth,trim=6 6 6 22.5,clip]
        {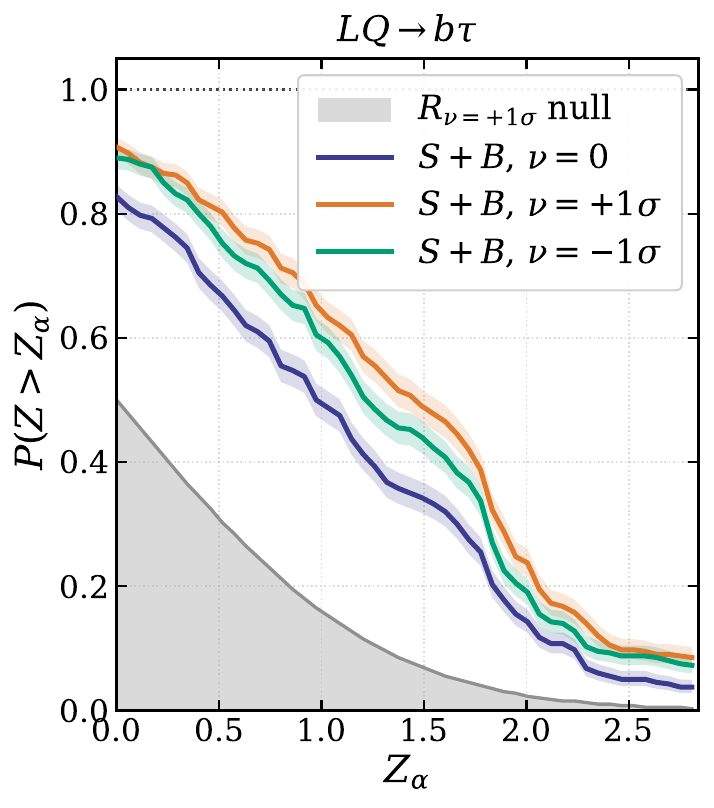}
        \caption{$LQ\to bt$}
    \end{subfigure}\hfill
    \begin{subfigure}[t]{0.247\textwidth}
        \centering
        \includegraphics[width=\linewidth,trim=6 6 6 22.5,clip]
        {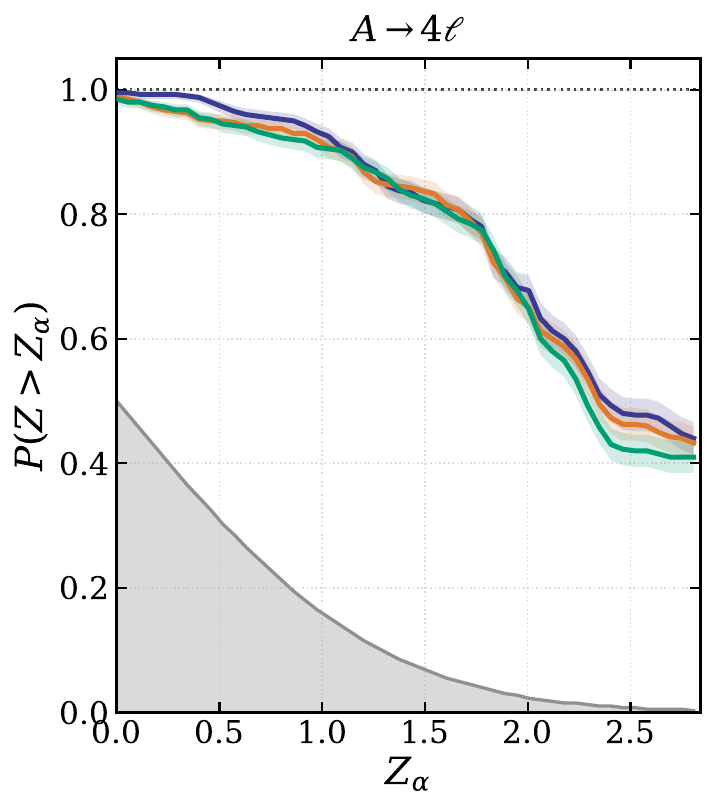}
        \caption{$A\to 4\ell$}
    \end{subfigure}\hfill
    \begin{subfigure}[t]{0.247\textwidth}
        \centering
        \includegraphics[width=\linewidth,trim=6 6 6 22.5,clip]
        {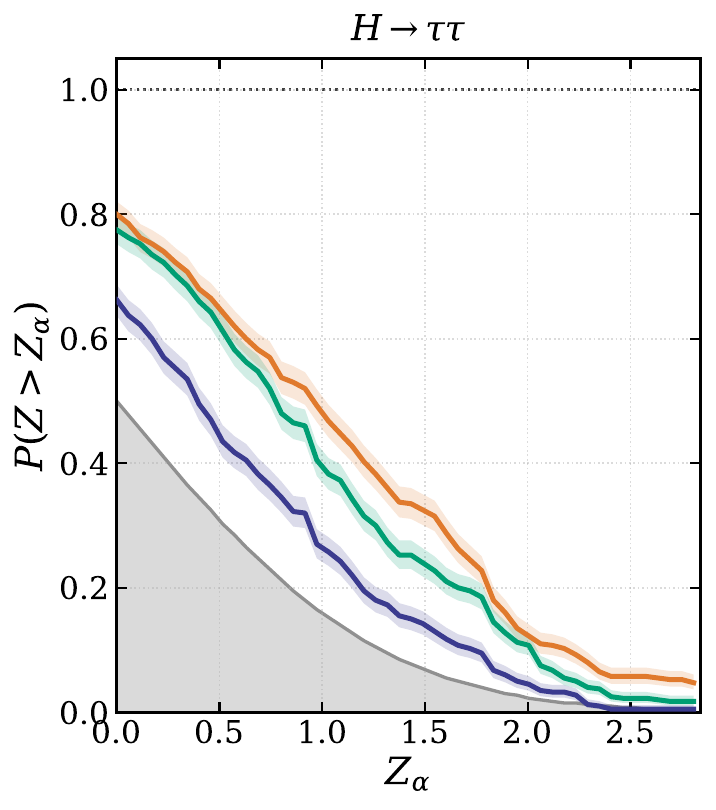}
        \caption{$h\to \tau\tau$}
    \end{subfigure}\hfill
    \begin{subfigure}[t]{0.247\textwidth}
        \centering
        \includegraphics[width=\linewidth,trim=6 6 6 22.5,clip]
        {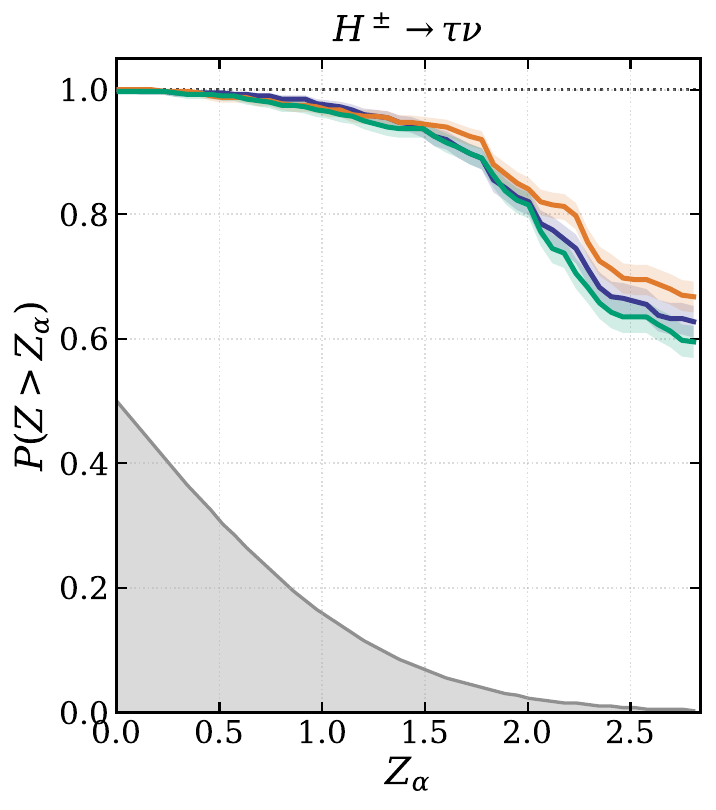}
        \caption{$H^\pm\to \tau^\pm\nu$}
    \end{subfigure}

    \par\vspace{0.6em}

    {\small\textbf{Linearized embedding}\par}
    \vspace{0.2em}

    \begin{subfigure}[t]{0.247\textwidth}
        \centering
        \includegraphics[width=\linewidth,trim=6 6 6 22.5,clip]
        {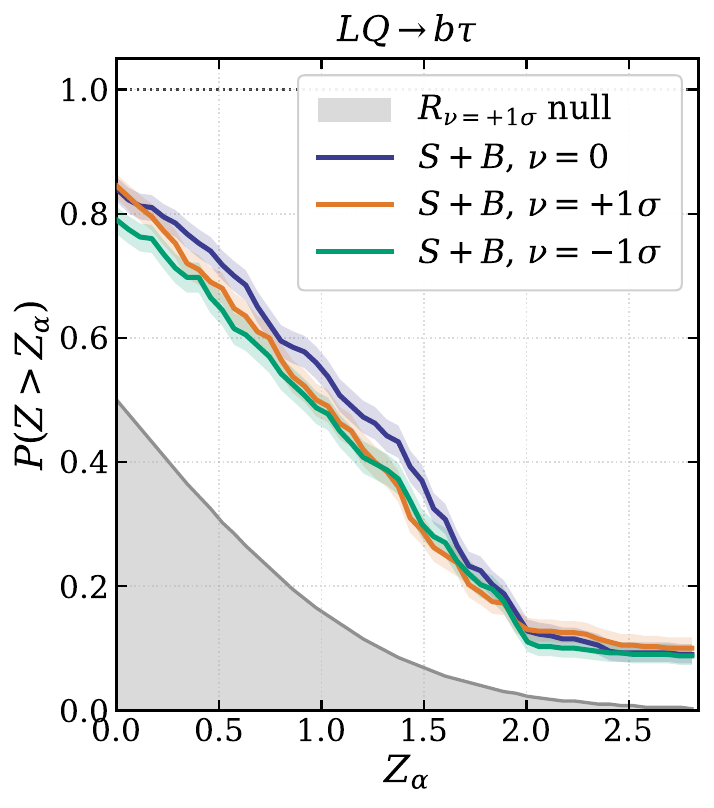}
        \caption{$LQ\to bt$}
    \end{subfigure}\hfill
    \begin{subfigure}[t]{0.247\textwidth}
        \centering
        \includegraphics[width=\linewidth,trim=6 6 6 22.5,clip]
        {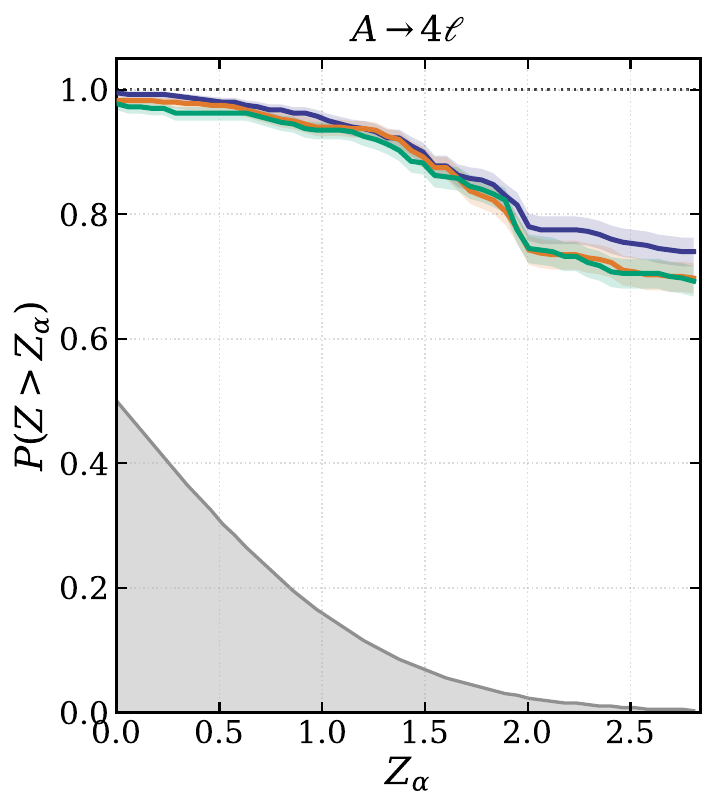}
        \caption{$A\to 4\ell$}
    \end{subfigure}\hfill
    \begin{subfigure}[t]{0.247\textwidth}
        \centering
        \includegraphics[width=\linewidth,trim=6 6 6 22.5,clip]
        {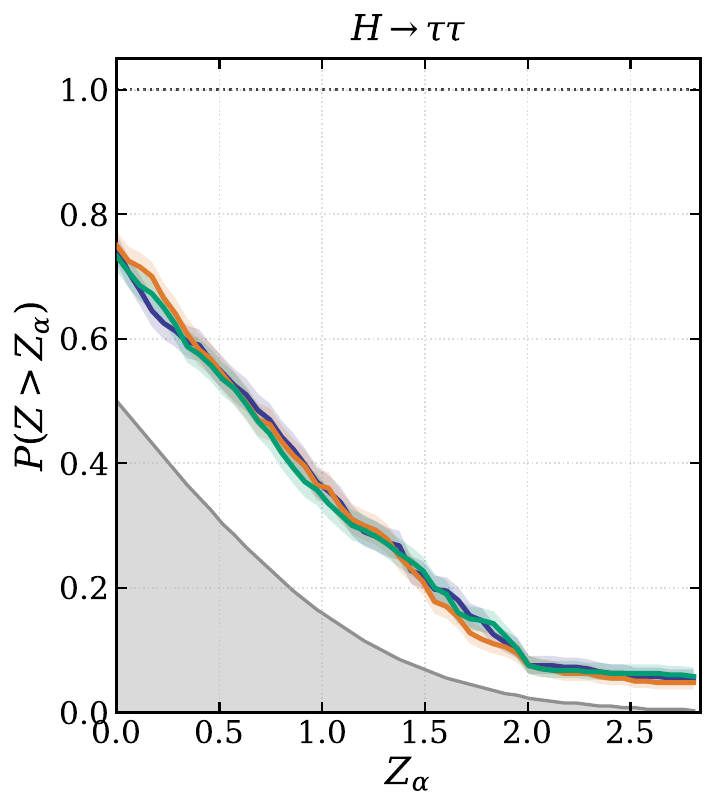}
        \caption{$h\to \tau\tau$}
    \end{subfigure}\hfill
    \begin{subfigure}[t]{0.247\textwidth}
        \centering
        \includegraphics[width=\linewidth,trim=6 6 6 22.5,clip]
        {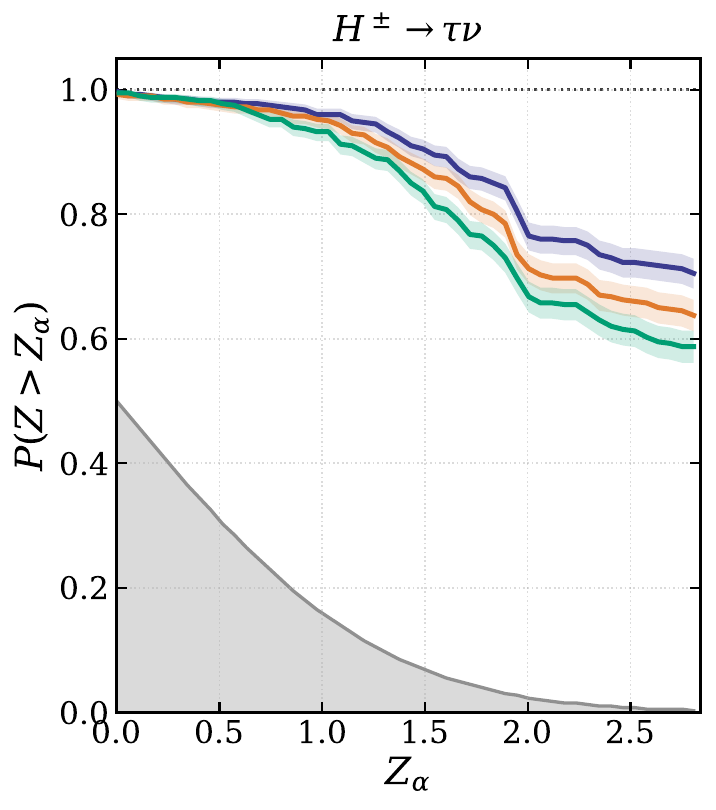}
        \caption{$H^\pm\to \tau^\pm\nu$}
    \end{subfigure}

    \caption{
    \textbf{Power curves for $1\%$ signal injection using the
    unlinearized and linearized embeddings.}
    The top row shows the unlinearized results, and the bottom row shows the linearized results, both empirically calibrated against $R_{\nu=+1\sigma}$. Each panel is evaluated using 400 null and 400 signal-injected toys. The signal benchmarks, nuisance settings,
    and gray reference curves follow the same conventions as in  Fig.~\ref{fig:power_curves_emp_n50}.}
    \label{fig:power_curves_emp_n100}
\end{figure*}

The power curves in Figures~\ref{fig:power_curves_emp_n50} and~\ref{fig:power_curves_emp_n100} provide a richer view of the discovery sensitivity, inspecting all discovery thresholds simultaneously for $0.5\%$ and $1\%$ signal injections, respectively. Each figure compares the unlinearized embedding in the top row with the linearized embedding in the bottom row, both calibrated against $R_{\nu=+1\sigma}$. Each curve gives the probability that the observed significance exceeds a threshold $Z_\alpha$; better performance is indicated by signal-injected curves that lie well above the gray calibration curve and retain high power as $Z_\alpha$ increases. For the linearized embedding, the power curves corresponding to different nuisance values overlap closely in most cases, indicating that the discovery power is both strong and mostly independent of $\nu$. In contrast, the unlinearized embedding shows much larger shifts and fluctuations across nuisance values. Overall, the linearized embedding yields power curves that are nearly independent of $\nu$ and, at the nominal value $\nu=0$, lie at or above those of the unlinearized embedding. The higher power of the unlinearized embedding at $\nu=\pm1\sigma$ in some panels reflects the lack of background closure rather than genuine discovery sensitivity. These trends are consistent with the test-statistic distributions in Fig.~\ref{fig:sensitivity} and the median $z$-scores reported in Tables~\ref{tab:linearized} and~\ref{tab:unlinearized}: the linearized embedding gives stable sensitivity across nuisance values, whereas the unlinearized embedding produces a less reliable and strongly nuisance-dependent response.
 
\subsection{Computational Resources.}
Encoder ($f_\theta$) training was performed on a 48 GB GPU, with a typical training time of 4 hours and 45 minutes. Linear model ($g_\phi$) fine-tuning was performed on a 48 GB GPU, with a typical training time of 3 hours. The test network $h_\zeta$ was trained on CPUs, with a typical training time of 50 minutes.

\subsection{Code Availability.}
The code and detailed documentation to reproduce the experiments can be found at \url{https://github.com/tshelley200/cl4ad_uncertainty}.

\section{Conclusion}
\label{sec:conclusion}

We introduced a method for incorporating continuous systematic uncertainties into contrastive latent spaces. The approach imposes a controlled linear covariance structure, with the goal of enabling statistically rigorous profile-likelihood anomaly detection at the LHC. 
We regularized the training of a contrastive neural embedding using an auxiliary objective that encourages the learned log-density ratio to vary linearly with the nuisance parameter. This keeps the latent space compact, while preserving the information needed to model systematic variations.

This construction integrates naturally with the anomaly detection framework of~\cite{dAgnolo:2021aun}. It provides a scalable workflow in which nuisance-dependent effects can be included directly in the likelihood-based test statistic, rather than treated through post-hoc calibration.

The results presented in this work demonstrate that we can build anomaly detection methods that capture high-dimensional event structure, while remaining tractable and enabling robust and interpretable inference. 
Contrastive representations can be made compatible with continuous nuisance profiling, allowing the quantitative study of systematic uncertainties affecting an arbitrarily complex input representation in the flat, low-dimensional representation of an embedded space. 

Enforcing linearity in $\nu$ carries two additional benefits: (1) it simplifies the modelling and combination of multiple nuisance parameters and (2) it enables powerful combination of the test statistic across batches of data~\cite{Grosso:2024nho}.
The proposed approach therefore offers a path toward the deployment of contrastive anomaly detection pipelines in full-scale LHC analyses, where a robust treatment of systematic uncertainties is essential.
Future work will extend the method to multiple nuisance parameters and to higher luminosity, towards more realistic experimental settings. 

\begin{acknowledgments}
This work is supported by the National Science Foundation under Cooperative Agreement PHY-2019786, the NSF AI Institute for Artificial Intelligence and Fundamental Interactions, \url{http://iaifi.org/}. Computations in this paper were run on the FASRC Cannon cluster supported by the FAS Division of Science Research Computing Group at Harvard University.
\end{acknowledgments}

\appendix

\section{Additional calibration tables}\label{app:closure}
We report in Tables~\ref{tab:validation_C} and \ref{tab:validation_B} the observed median z-scores of the signal injection experiments using as empirical calibration the test statistic distribution under $\nu=0$ and $\nu=-\sigma$. Results in these Tables are compatible with the ones reported in Table~\ref{tab:validation_D}, further confirming that the NPLM test was implemented in a robust way.

\begin{table*}
\centering
\small
\setlength{\tabcolsep}{5.5pt}
\renewcommand{\arraystretch}{1.6}

\caption{
\textbf{Signal-injection $z$-scores calibrated against the nominal background hypothesis $R_{0}$.}
Median observed $z$-scores for each injected signal benchmark when calibrated against the nominal empirical distribution, $R_{0}$. Uncertainties denote 68\% Clopper--Pearson confidence intervals. Values reported as $\geq 2.81$ saturate the finite-sample calibration limit.
}
\label{tab:validation_B}
\vspace{0.3em}

\begin{tabular}{c|cc|cc|cc|cc}
\hline

Test sample
& \multicolumn{2}{c|}{$LQ\to bt$}
& \multicolumn{2}{c|}{$A\to 4\ell$}
& \multicolumn{2}{c|}{$h\to \tau\tau$}
& \multicolumn{2}{c}{$H^\pm\to \tau^\pm\nu$} \\
\cline{2-9}
\footnotesize{(calibrated by $R_0$)}
& $0.5\%$ & $1\%$
& $0.5\%$ & $1\%$
& $0.5\%$ & $1\%$
& $0.5\%$ & $1\%$ \\
\hline
$\nu=0$
& $0.33^{+0.07}_{-0.06}$ & $1.05^{+0.09}_{-0.07}$
& $0.73^{+0.08}_{-0.07}$ & $\geq 2.81$
& $0.12^{+0.07}_{-0.06}$ & $0.60^{+0.08}_{-0.07}$
& $0.96^{+0.09}_{-0.07}$ & $\geq 2.81$ \\

$\nu=+1\sigma$
& $0.31^{+0.07}_{-0.06}$ & $0.91^{+0.08}_{-0.07}$
& $0.83^{+0.08}_{-0.07}$ & $\geq 2.81$
& $0.18^{+0.07}_{-0.06}$ & $0.56^{+0.07}_{-0.06}$
& $0.92^{+0.09}_{-0.07}$ & $\geq 2.81$ \\

$\nu=-1\sigma$
& $0.26^{+0.07}_{-0.06}$ & $0.86^{+0.08}_{-0.07}$
& $0.81^{+0.08}_{-0.07}$ & $\geq 2.81$
& $0.18^{+0.07}_{-0.06}$ & $0.56^{+0.07}_{-0.06}$
& $0.80^{+0.08}_{-0.07}$ & $\geq 2.81$ \\
\hline
\end{tabular}
\end{table*}

\begin{table*}
\centering
\small
\setlength{\tabcolsep}{5.5pt}
\renewcommand{\arraystretch}{1.6}

\caption{
\textbf{Signal-injection $z$-scores calibrated against $R_{\nu=-1\sigma}$.}
Median observed $z$-scores for each injected signal benchmark when calibrated against the empirical distribution of $R_{\nu=-1\sigma}$. Uncertainties denote 68\% Clopper--Pearson confidence intervals. Values reported as $\geq 2.81$ saturate the finite-sample calibration limit.
}
\label{tab:validation_D}

\vspace{0.3em}

\begin{tabular}{c|cc|cc|cc|cc}
\hline
Test sample
& \multicolumn{2}{c|}{$LQ\to bt$}
& \multicolumn{2}{c|}{$A\to 4\ell$}
& \multicolumn{2}{c|}{$h\to \tau\tau$}
& \multicolumn{2}{c}{$H^\pm\to \tau^\pm\nu$} \\
\cline{2-9}
\footnotesize{(calibrated by $R_{\nu=-1\sigma}$)}
& $0.5\%$ & $1\%$
& $0.5\%$ & $1\%$
& $0.5\%$ & $1\%$
& $0.5\%$ & $1\%$ \\
\hline

$\nu=0$
& $0.28^{+0.07}_{-0.06}$ & $1.14^{+0.10}_{-0.08}$
& $0.84^{+0.08}_{-0.07}$ & $\geq 2.81$
& $0.09^{+0.07}_{-0.06}$ & $0.61^{+0.08}_{-0.07}$
& $0.99^{+0.09}_{-0.07}$ & $\geq 2.81$ \\

$\nu=+1\sigma$
& $0.28^{+0.07}_{-0.06}$ & $0.97^{+0.09}_{-0.07}$
& $0.96^{+0.09}_{-0.07}$ &$\geq 2.81$
& $0.13^{+0.07}_{-0.06}$ & $0.57^{+0.07}_{-0.06}$
& $0.97^{+0.09}_{-0.07}$ & $\geq 2.81$ \\

$\nu=-1\sigma$
& $0.23^{+0.07}_{-0.06}$ & $0.97^{+0.09}_{-0.07}$
& $0.92^{+0.09}_{-0.07}$ & $\geq 2.81$
& $0.14^{+0.07}_{-0.06}$ & $0.56^{+0.07}_{-0.06}$
& $0.91^{+0.08}_{-0.07}$ & $\geq 2.81$ \\

\hline
\end{tabular}
\end{table*}
We report in Table~\ref{tab:validation_E} the observed median z-scores of the signal injection experiments using as calibration the asymptotic $\chi^2_{45}$. The results in this Table are compatible with the ones reported in Table~\ref{tab:validation_B}, further confirming the compatibility of the null distribution with $\chi^2_{45}$ obtained via weight clipping tuning.
\begin{table}[h]
\centering
\small
\setlength{\tabcolsep}{2pt}
\renewcommand{\arraystretch}{1.6}

\caption{
\textbf{Signal-injection $z$-scores from analytical $\chi^2_{45}$ calibration.}
$z$-scores obtained using the analytical $\chi^2$ calibration with 45 degrees of freedom. Each signal benchmark is evaluated at $0.5\%$ and $1\%$ signal injection, shown in the left and right columns, respectively. For each benchmark and injection fraction, we generate 400 null toys and 400 signal-injected toys.}
\label{tab:validation_E}
\begin{tabular}{c|cc|cc|cc|cc}
\hline
Test sample
& \multicolumn{2}{c|}{$LQ\to bt$}
& \multicolumn{2}{c|}{$A\to 4\ell$}
& \multicolumn{2}{c|}{$h\to \tau\tau$}
& \multicolumn{2}{c}{$H^\pm\to \tau^\pm\nu$} \\
\cline{2-9}
\footnotesize{(cal. by $\chi^2_{45}$)}
& $0.5\%$ & $1\%$
& $0.5\%$ & $1\%$
& $0.5\%$ & $1\%$
& $0.5\%$ & $1\%$ \\
\hline

$\nu=0$
& 0.33 & 1.08
& 0.82 & 3.19
& 0.17 & 0.65
& 0.99 & 3.05 \\

$\nu=+1\sigma$
& 0.31 & 0.96
& 0.90 & 3.03
& 0.20 & 0.61
& 0.96 & 2.80 \\

$\nu=-1\sigma$
& 0.28 & 0.93
& 0.87 & 3.06
& 0.21 & 0.61
& 0.86 & 2.72 \\
\hline
\end{tabular}
\end{table}

\section{The effect of linearization in the latent representation}
Table~\ref{tab:embedding_linearization_no_nuisance} shows the detection performance of NPLM in the absence of uncertainties, for both the original embedding, trained with supervised contrastive learning alone, and our proposed embedding, co-trained with a linearization penalty. While this is not a realistic scenario it informs us about the properties of the embedding as an input to the NPLM test.
After linearization the sensitivity of NPLM is moderately degraded, telling us that accounting for systematics in the embedding come at the cost of degraded sensitivity. This is a price we are willing to pay in order to make the detection robust and reliable.
\begin{table*}[!]
\centering
\small
\setlength{\tabcolsep}{4.5pt}
\renewcommand{\arraystretch}{1.6}
\caption{
\textbf{Comparison of idealized signal-injection $z$-scores before and after embedding linearization, without nuisance parameters included in the NPLM test.}
Median observed $z$-scores for each injected signal benchmark, evaluated using nominal samples with $\nu=0$ and calibrated against the empirical nominal
background-only distribution $R_0$. Each signal benchmark is evaluated at $0.5\%$ and $1\%$ signal injection, shown in the left and right columns, respectively. For each benchmark and injection fraction, we generate 400 null toys and 400 signal-injected toys. Comparing the two rows isolates the effect
of embedding linearization in the absence of systematic variations and nuisance parameters. This
table represents an idealized sensitivity that is not
achievable in a realistic analysis, where uncertainties cannot be ignored and
must instead be incorporated as nuisance parameters in the NPLM test.
Uncertainties denote 68\% Clopper--Pearson confidence intervals. Values reported
as $\geq 2.81$ saturate the finite-sample calibration limit.
}
\label{tab:embedding_linearization_no_nuisance}
\vspace{0.3em}

\begin{tabular}{c|cc|cc|cc|cc}
\hline
\multirow{2}{*}{Embedding}
& \multicolumn{2}{c|}{$LQ\to bt$}
& \multicolumn{2}{c|}{$A\to 4\ell$}
& \multicolumn{2}{c|}{$h\to\tau\tau$}
& \multicolumn{2}{c}{$H^\pm\to\tau^\pm\nu$} \\
\cline{2-9}
& $0.5\%$ & $1\%$
& $0.5\%$ & $1\%$
& $0.5\%$ & $1\%$
& $0.5\%$ & $1\%$ \\
\hline

Before Linearization
& $0.38^{+0.07}_{-0.06}$ & $1.70^{+0.15}_{-0.10}$
& $1.01^{+0.09}_{-0.07}$ & $\geq 2.81$
& $0.29^{+0.07}_{-0.06}$ & $1.13^{+0.10}_{-0.08}$
& $1.42^{+0.12}_{-0.09}$ & $\geq 2.81$ \\

After Linearization
& $0.37^{+0.07}_{-0.06}$ & $1.21^{+0.10}_{-0.08}$
& $1.06^{+0.09}_{-0.07}$ & $\geq 2.81$
& $0.17^{+0.07}_{-0.06}$ & $0.76^{+0.08}_{-0.07}$
& $0.96^{+0.09}_{-0.07}$ & $\geq 2.81$ \\
\hline
\end{tabular}
\end{table*}

\section{Quadratic Nuisance Model}\label{app:quadratic}

Throughout the main text, we use a linear nuisance model in the NPLM test, corresponding to the simplest nontrivial polynomial dependence on the nuisance parameter. This choice is motivated primarily by its use in ``stacking,'' where multiple datasets are combined to increase the effective luminosity beyond that of the individual 10,000-event samples considered here; a detailed treatment of this procedure is left to a separate future work. However, a higher-order polynomial model can generally better describe residual nonlinearities in the nuisance dependence across the embedding space. In particular, a quadratic model can improve closure where the linear approximation is insufficient and yield more stable signal significances across different nuisance values.

In this section, we evaluate a quadratic nuisance model using the same linearized embeddings, closure tests, and signal-injection studies as in the main text. In Table~\ref{tab:validation_quadratic}, we report compatibility tests between the nominal and nuisance-shifted background-only distributions, $R_0$ and $R_\nu$, and in Table~\ref{tab:validation_quadratic_sig} the observed median $Z$-scores for the signal-injection experiments, calibrated against the empirical nominal distribution $R_0$. The results are broadly consistent with those obtained using the linear nuisance model in Tables~\ref{tab:validation_B} and~\ref{tab:validation_C}, with slight improvements in closure tests. The quadratic model yields results in close agreement with those of the linear model, indicating that the additional complexity provides no meaningful improvement. The linear assumption is therefore good enough for this study.

\begin{table}[!tbp]
\centering
\small
\setlength{\tabcolsep}{8pt}
\renewcommand{\arraystretch}{1.4}

\caption{
\textbf{Closure validation of the NPLM test using a quadratic nuisance model.}
The table reports $p$-values comparing each nuisance-shifted
background-only test-statistic distribution, $R_{\nu=\pm1\sigma}$,
with the empirical nominal distribution $R_0$.
Each distribution contains 400 replicas, with nuisance parameters
included in the NPLM test and the embedding evaluated at the selected
weight-clipping value $w_{\mathrm{clip}}=1.94$.
Closure is assessed using the two-sample Kolmogorov--Smirnov test
and binned Pearson $\chi^2$ tests with the degrees of freedom
listed below.
}
\label{tab:validation_quadratic}

\begin{tabular}{lccc}
\hline
Test & DoF & $\nu=+1\sigma$ & $\nu=-1\sigma$ \\
\hline
Kolmogorov--Smirnov
    & --- & $0.984$  & $0.813$  \\
\hline
Pearson $\chi^2$
    & 10  & $0.8082$ & $0.7929$ \\
    & 14  & $0.9786$ & $0.6844$ \\
    & 17  & $0.9797$ & $0.7538$ \\
    & 20  & $0.9793$ & $0.5526$ \\
    & 25  & $0.9305$ & $0.8138$ \\
\hline
\end{tabular}
\end{table}

\begin{table*}[!]
\centering
\small
\setlength{\tabcolsep}{5.5pt}
\renewcommand{\arraystretch}{1.6}
\caption{
\textbf{Signal-injection $z$-scores with nuisance parameters included in the NPLM test.}
Median observed $z$-scores for each injected signal benchmark and test-sample nuisance value, calibrated against the empirical nominal background-only distribution $R_0$. Each signal benchmark is evaluated at $0.5\%$ and $1\%$ signal injection, shown in the left and right columns, respectively. For each benchmark and injection fraction, we generate 400 null toys and 400 signal-injected toys. Uncertainties denote 68\% Clopper--Pearson confidence
intervals. Values reported as $\geq 2.81$ saturate the finite-sample calibration limit.}
\label{tab:validation_quadratic_sig}
\vspace{0.3em}

\begin{tabular}{c|cc|cc|cc|cc}
\hline
\multirow{2}{*}{Test sample}
& \multicolumn{2}{c|}{$LQ\to bt$}
& \multicolumn{2}{c|}{$A\to 4\ell$}
& \multicolumn{2}{c|}{$h\to \tau\tau$}
& \multicolumn{2}{c}{$H^\pm\to \tau^\pm\nu$} \\
\cline{2-9}
& $0.5\%$ & $1\%$
& $0.5\%$ & $1\%$
& $0.5\%$ & $1\%$
& $0.5\%$ & $1\%$ \\
\hline
$\nu=0$
& $0.28^{+0.07}_{-0.06}$ & $1.13^{+0.10}_{-0.08}$
& $0.90^{+0.08}_{-0.07}$ & $\geq 2.81$
& $0.25^{+0.07}_{-0.06}$ & $0.71^{+0.08}_{-0.07}$
& $0.98^{+0.09}_{-0.07}$ & $\geq 2.81$ \\

$\nu=+1\sigma$
& $0.27^{+0.07}_{-0.06}$ & $0.83^{+0.08}_{-0.07}$
& $0.81^{+0.08}_{-0.07}$ & $\geq 2.81$
& $0.10^{+0.07}_{-0.06}$ & $0.53^{+0.07}_{-0.06}$
& $0.81^{+0.08}_{-0.07}$ & $\geq 2.81$ \\

$\nu=-1\sigma$
& $0.27^{+0.07}_{-0.06}$ & $0.88^{+0.08}_{-0.07}$
& $0.81^{+0.08}_{-0.07}$ & $\geq 2.81$
& $0.12^{+0.07}_{-0.06}$ & $0.47^{+0.07}_{-0.06}$
& $0.88^{+0.08}_{-0.07}$ & $2.43^{+0.48}_{-0.16}$ \\
\hline
\end{tabular}
\end{table*}

\clearpage
\nocite{*}

\bibliography{apssamp}

\end{document}